\documentclass{emulateapj}

\newcommand{\Angstrom}{${\buildrel _{\circ} \over{\mathrm{A}}} \;$}

\slugcomment{Accepted for publication in The Astrophysical Journal.}
\shorttitle{Panchromatic Estimation of SFRs}
\shortauthors{Kurczynski et al.}

\usepackage{graphicx}

\DeclareGraphicsExtensions{.pdf,.png,.jpg,.eps}

\usepackage{amssymb, amsmath}

\usepackage[hyperindex,breaklinks]{hyperref}

\begin{document}


\title{Panchromatic Estimation of Star Formation Rates in {\it BzK} Galaxies at 1$<$z$<$3}



\author{Peter Kurczynski\altaffilmark{1}, Eric Gawiser\altaffilmark{1}, Minh Huynh\altaffilmark{2,3}, Rob J. Ivison\altaffilmark{4}, Ezequiel Treister\altaffilmark{5,6}, Ian Smail\altaffilmark{7}, Guillermo A. Blanc\altaffilmark{8}, Carolin N. Cardamone\altaffilmark{9}, Thomas R. Greve\altaffilmark{10}, Eva Schinnerer\altaffilmark{10}, Meg Urry\altaffilmark{11}, Paul van der Werf\altaffilmark{12}}

\altaffiltext{1}{Department of Physics and Astronomy, Rutgers University, Piscataway, NJ 08854, USA}
\altaffiltext{2}{International Centre for Radio Astronomy Research, University of Western Australia M468, 35 Stirling Highway, Crawley WA 6009, Australia}
\altaffiltext{3}{Infrared Processing and Analysis Center, MS 220-6, California Institute of Technology, Pasadena, CA 91125, USA}
\altaffiltext{4}{UK Astronomy Technology Centre, Royal Observatory, Blackford Hill, Edinburgh EH9 3HJ, UK}
\altaffiltext{5}{Institute for Astronomy, University of Hawaii, 2680 Woodlawn Drive, Honolulu, HI 96822, USA.}
\altaffiltext{6}{Departamento de Astronom\'{\i}a, Universidad de Concepci\'{o}n, Casilla 160-C, Concepci\'{o}n, Chile.}
\altaffiltext{7}{Institute for Computational Cosmology, Department of Physics, Durham University, South Road, Durham DH1 3LE, UK}
\altaffiltext{8}{Department of Astronomy, The University of Texas at Austin, Austin, TX, USA}
\altaffiltext{9}{Physics Department, Massachusetts Institute of Technology, Cambridge, MA 02139, USA}
\altaffiltext{10}{MPI for Astronomy, K\"{o}nigstuhl 17, 69117 Heidelberg, Germany}
\altaffiltext{11}{Department of Physics, Yale University, P.O. Box 208121, New Haven, CT 06520-8121, USA}
\altaffiltext{12}{Leiden Observatory, Leiden University, P.O. Box 9513, NL - 2300 RA Leiden, The Netherlands}




\begin{abstract}

We determine star formation rates (SFRs) in a sample of color-selected, star-forming (s{\it BzK}) galaxies ($K_{AB}<21.8$) in the Extended {\it Chandra} Deep Field - South.  To identify and avoid active galactic nuclei, we use X-ray, IRAC color, and IR/radio flux ratio selection methods.  Photometric redshift-binned, average flux densities are measured with stacking analyses in {\it Spitzer}-MIPS IR, BLAST and APEX/LABOCA submillimeter, VLA and GMRT radio and {\it Chandra} X-ray data.  We include averages of aperture fluxes in MUSYC $UBVRIz'JHK$ images to determine UV-through-radio spectral energy distributions.  We determine the total IR luminosities and compare SFR calibrations from FIR, 24 $\mu$m, UV, radio and X-ray wavebands.  We find consistency with our best estimator, SFR$_{IR+UV}$, to within errors for the preferred radio SFR calibration.  Our results imply that 24 $\mu$m only and X-ray SFR estimates should be applied to high redshift galaxies with caution.  Average IR luminosities are consistent with luminous infrared galaxies.  We find SFR$_{IR+UV}$ for our stacked s{\it BzK}s at median redshifts 1.4, 1.8, and 2.2 to be 55$\pm$6 (random error), 74$\pm8$ and 154$\pm17$ M$_\odot$ yr$^{-1}$ respectively, with additional systematic uncertainty of a factor of $\sim$~2. 
\end{abstract}


\section{INTRODUCTION}
\label{IntroductionSection}

The history of star formation traces the origins of visible matter in the universe.  Understanding star formation across cosmic time will yield insights into diverse areas of astronomy from the formation and evolution of galaxies to the initial conditions of stellar evolution.  The redshift range $1<z<3$ is a key epoch in this history, when most of the stars in the universe were born.  Galaxies in this range are identified with color selection methods.  Estimating their Star Formation Rates (SFRs) is complicated by contamination from Active Galactic Nuclei (AGN), and differences in luminosity to SFR calibration in the various wavebands used.\footnote{Other sources of systematic uncertainty include the initial mass function (IMF), photometric redshifts, and spectral energy distributions (SEDs).}

SFRs are estimated from a wide variety of luminosity calibrations from X-ray through radio wavebands.  Broadband UV continuum radiation directly probes the light of young stars, but is strongly attenuated by dust.  The thermal IR luminosity, hereafter defined as L$_{IR}$,  L$_{IR} \equiv L(8-1000~\mu m)$, measures the SFR from the reprocessed dust emission (e.g. \citealt{1998ApJ...498..541K}).  IR and uncorrected UV luminosities represent reprocessed and un-reprocessed photons and are used to estimate the total SFR.  

UV slope based corrections for dust attenuation (e.g. \citealt{1999ApJ...521...64M}) have been previously applied to high redshift, star-forming galaxies (e.g. \citealt{2004ApJ...603L..13R,2005ApJ...633..748R,2006ApJ...644..792R,2010ApJ...712.1070R,2007ApJ...670..156D,2010ApJ...714.1740M}).  Low redshift analogs of Lyman Break Galaxies (LBGs) have similar dust attenuation corrections \citep{2041-8205-726-1-L7}, although ULIRGs have higher dust correction factors \citep{2010ApJ...715..572H} and other star-forming galaxies have lower dust corrections than expected (e.g.\citealt{2010MNRAS.409L...1B}), and significant scatter is observed.    

Radio-wave (1.4 GHz) luminosity in star-forming galaxies, primarily synchrotron emission from supernova remnants, also traces SFR \citep{1992ARA&A..30..575C}.  Radio-wave SFR calibrations rely explicitly upon the IR-radio correlation (e.g. \citealt{2003ApJ...586..794B,2001ApJ...554..803Y}) or use an implicit conversion to H$\alpha$ luminosity to calibrate SFR  \citep{1992ARA&A..30..575C}.

X-ray emission in star-forming galaxies arises from low mass X-ray binaries (LMXBs), consisting of long lived, low mass (M$< 1 M_\odot$) stars with neutron star companions, high mass X-ray binaries (HMXBs), consisting of short-lived, massive (M$>8 M_\odot$) stars with a neutron star companion, and to a lesser extent, supernova remnants \citep{2002A&A...382..843P}.   These last two sources of X-ray emission are linked to short-lived, massive stars, providing a rationale for X-ray SFR calibrations.   However, these multiple sources of X-ray emission, as well as X-ray obscuration by gas and dust, complicate X-ray luminosity to SFR calibrations.  In practice, X-ray SFR calibrations (e.g. \citealt{2003A&A...399...39R,2004A&A...419..849P,2010ApJ...724..559L}) explicitly rely upon empirical correlations with IR luminosity and are therefore indirect measures of IR luminosity.   X-ray emission from AGN are major sources of contamination.
  
\citet{2007ApJ...670..156D}
find SFR  from dust-corrected UV, 24 $\mu$m, 1.4 GHz, and X-ray calibrations to be approximately consistent for star-forming ({\it BzK}; see below) galaxies at $z\sim2$.  \citet{2006ApJ...644..792R} compare dust-corrected UV, 24 $\mu$m and X-ray calibrations for a spectroscopic sample of $z\sim2$ galaxies, and find dust-corrected UV SFR to be consistent with 24 $\mu$m SFR for {\it BzKs}, but not other types of star-forming galaxies.

\citet{2011ApJ...738..106W} compare dust-corrected UV, 24 $\mu$m and {\it Herschel} PACS derived SED based SFRs to H$\alpha$ based SFRs and find H$\alpha$ derived SFRs to require extra dust correction to agree with SFR$_{UV+IR}$.  In addition, other recent literature has illustrated that 24 $\mu$m SFRs are overestimated for high luminosity sources at $z\sim2$ \citep{2011A&A...533A.119E,2012ApJ...745..182N}.

\citet{2009ApJ...698L.116P} find radio SFR based on the calibration of \citet{2001ApJ...554..803Y}
to be consistent with dust-corrected UV SFR for {\it BzKs}.

X-ray SFRs based on the calibration of \citet{2003A&A...399...39R} are found to agree with dust-corrected UV SFRs for BX and BM galaxies in the range $1.5 < z \leq 3.0$ \citep{2004ApJ...603L..13R} and with 24 $\mu$m SFR \citep{2006ApJ...644..792R}.

We seek to expand upon previous studies by including a wider range of luminosity estimates, including the most recent results from X-ray observations and analyses, and by understanding the assumptions and sources of uncertainty inherent to each SFR calibration.  We seek to ascertain the extent to which submillimeter data can improve IR-based SFR estimates.  We compare different radio luminosity to SFR calibrations, and we use the most recent X-ray data and luminosity to SFR calibrations. By comparing these estimates to the total SFR, derived from the sum of IR and uncorrected UV luminosity, we seek to test their robustness at high redshift.

We investigate SFRs binned according to photometric redshift, with stacking analyses in radio through X-ray wavebands.  In particular, we use extensive FIR-submillimeter data, for which we have an improved stacking algorithm \citep{2010AJ....139.1592K}.  We discuss observations and data in Section \ref{DataSection}.  We present our stacking methodology in Section \ref{MethodSection}.   Results are presented in Section \ref{ResultsSection}, and we discuss comparisons of SFR in Section \ref{DiscussionSection}.  Our conclusions are summarized in Section \ref{ConclusionSection}.  Throughout this paper, magnitudes are measured in the AB system unless stated otherwise.  We assume a \citet{1955ApJ...121..161S} IMF from 0.1 to 100 M$_\odot$.  Another commonly used IMF, that of \citet{2001ASPC..228..187K}, would change the slope of the low mass end of the IMF, and multiply the SFRs presented here by factors of 0.58 (e.g. \citealt{2007ASPC..380..423H}).  We adopt a cosmology with $\Omega_\Lambda$ = 0.7, $\Omega_0$ = 0.3, and h$_{100}$ = 0.7.
%
\vspace{0.1in}
\section{SAMPLE SELECTION}
\label{DataSection}
The {\it BzK} color selection criterion has emerged as a successful color based method for identifying galaxies in the range $1.4\lesssim z \lesssim 2.5$ in a maximally inclusive manner  \citep{2004ApJ...617..746D}. The quantity {\it BzK}\footnote{In this paper, we have corrected colors to account for differences between the Bessel $B$ band filter used on the Very Large Telescope (VLT) in the original sample of  \citet{2004ApJ...617..746D} and the Johnson $B$ band filter on WFI, Suprime-Cam and other instruments.  Not accounting for this correction can lead to an offset of 0.5 mag toward lower values in ($B$ - z) color and 0.04 mag in higher (z - $K$) color, producing a significant excess of ``s{\it BzK}'' galaxies that are in fact low redshift contaminants \citep{2008ApJ...681.1099B}.} is defined as
\begin{equation}
BzK \equiv (z-K)_{AB} - (B-z)_{AB}
\end{equation}
Actively star-forming galaxies, s{\it BzK}s, are found to satisfy {\it BzK}$ >$ -0.2, the upper left region in Figure \ref{FIG:BzKDiagram}.  The reddening vector in the {\it BzK} plane is parallel to the {\it BzK} line, making this selection unbiased with respect to dust content.  In this paper, we ignore the reddest galaxies in both z-$K$ and $B$-z, p{\it BzK}s, which tend to be old, passively evolving stellar systems, and are located in the upper right region of Figure \ref{FIG:BzKDiagram}.  Comparisons and overlaps between {\it BzK}s and other color-selected galaxy types are discussed in \citet{2005ApJ...633..748R}, \citet{2007A&A...465..393G}, and \citet{2010ApJ...719..483G}.
SFR estimates typically range from several tens to hundreds M$_\odot$ yr$^{-1}$ (e.g.  \citealt{2004ApJ...617..746D,2005ApJ...631L..13D,2005ApJ...633..748R,2005ApJ...631L..13D,2007ApJ...670..156D,2006ApJ...644..792R,2009MNRAS.394....3D,2009ApJ...698L.116P,2010ApJ...719..483G,2010ApJ...718..112Y}). 

Our sample  of s{\it BzK} galaxies with $K_{Vega} < $ 20 ($K_{AB} < $ 21.8) comes from the catalog of \citet{2008ApJ...681.1099B}, and is taken from the Multiwavelength Survey by Yale-Chile (MUSYC; \citealt{2006ApJS..162....1G}) observations of the Extended {\it Chandra} Deep Field - South (ECDF-S).  Photometry in the $UBVRIz'JHK$ wavebands is obtained from the $K$-selected catalog of  \citet{2009ApJS..183..295T}, and is augmented with data from the MUSYC optical catalog ($R_{AB} < 25.3$ depth; \citealt{2010ApJS..189..270C}), which includes photometry from 32 bands including 18 medium band optical filters and {\it {\it Spitzer}} Infrared Array Camera (IRAC) bands.  

We use X-ray luminosity, IRAC colors and q$_{IR}$ criteria to discriminate SF galaxies from AGN.   There are 110 {\it BzK} sources detected in the combined 250 ks \citep{2006AJ....131.2373V} + 4 Ms \citep{2011ApJS..195...10X} {\it Chandra} catalogs.  Of these sources, 61 are in the {\it Chandra} Deep Field - South (CDF-S), and the remaining 49 are in the ECDF-S.  X-ray luminosity is used to distinguish AGN from SF galaxies 
(e.g. \citealt{2002ApJ...576..625N}):
\begin{equation}
L(\text{AGN; 2-10~keV}) > 10^{42} \text{erg~s}^{-1}
\end{equation}
There are 107/110 X-ray detected {\it BzK}s that meet this criterion (three X-ray detected {\it BzK}s in the 4 Ms CDF-S have L$_X < 10^{42}$ erg s$^{-1}$ and are considered SF galaxies in this analysis).  25 {\it BzK}s were identified as AGN on the basis of their position in IRAC color-color space \citep{0004-637X-748-2-142}.
Finally, we computed q$_{IR}$ values, defined as the ratio of integrated IR flux, FIR (8-1000 $\mu$m; rest frame; W m$^{-2}$) to radio flux density, F$_{1.4~GHz}$ (rest frame; W m$^{-2}$ Hz$^{-1}$) 
(e.g. \citealt{2010A&A...518L..31I}):
\begin{equation}
q_{IR} = \log(\text{FIR} / 3.75\times10^{12}~\text{Hz} / F_{1.4~\text{GHz}} )  
\end{equation}

Radio loud AGN are discriminated from radio-quiet AGN / SF galaxies according to 
 \citet{2010A&A...518L..31I}:
\begin{equation}
q_{IR} (\text{AGN}) <  2.0  
\end{equation}

Computing q$_{IR}$ for {\it BzK}s requires sufficient photometry to estimate the rest-frame FIR SED.  Observed frame SEDs for {\it BzK}s with IRAC+MIPS24+VLA detections were fit  to \citealt{2001ApJ...556..562C} (hereafter CE01)
templates.  Rest-frame radio luminosities were estimated by assuming an $S_\nu \sim \nu^\alpha$, $\alpha = -0.8$ radio SED, characteristic of SF galaxies \citep{1992ARA&A..30..575C}.  Thirty-five sources were identified as AGN using the q$_{IR}$ criterion; 15 of these sources are uniquely identified as AGN with this method, and the remainder are also identified as AGN using other methods.  

A summary of the results of AGN rejection methods is given in Table \ref{TABLE:AGNCRITERIA}.  Sources identified as AGN are excluded from the remaining analysis.  At present it is not possible to conclusively identify all AGN including deeply obscured sources; however, these methods make the best use of available data.  Effects of possible residual AGN contamination are discussed in Section \ref{DiscussionSection}.

Photometric or spectroscopic redshifts are assigned to s{\it BzK} galaxies by matching positions in the $K$ band source catalog with the MUSYC optical catalog and using redshifts from the optical catalog where redshifts in both catalogs are available.  Photometric redshifts for the optical catalog were obtained with the EAzY software \citep{2008ApJ...686.1503B}; the reader is referred to  \citet{2010ApJS..189..270C} for details of the redshift determination.

Histograms of photometric and spectroscopic redshifts are illustrated in Figure \ref{FIG:ZHistograms}.  These histograms indicate that the majority of s{\it BzK}s fall in the traditional redshift range associated with {\it BzK} galaxies, $1.4<z<2.5$. There are significant numbers of s{\it BzK}s at lower (20\%)and higher (4\%) redshifts, although {\it BzK} selection is less efficient in these ranges.  Spectroscopic redshifts are taken from the literature, for which there are substantial selection biases; therefore we do not expect the distribution of spectroscopic redshifts to match precisely the distribution of photometric redshifts.  Thus, the trend of spectroscopic redshifts being distributed at somewhat higher redshifts than the photometric redshifts, apparent from Figure \ref{FIG:ZHistograms}, is not an indication of systematic photometric redshift error.  

There are 48 galaxies in the range $1.2<z<3.3$ that have spectroscopic and photometric redshifts available.  Photometric redshift errors, defined as $(z_{phot} - z_{spec})/(1+z_{spec})$, are well described by a Gaussian with mean = -0.02 and $\sigma$=0.09, see Figure \ref{FIG:PhotozVsSpecz}.  Photometric redshift errors for these galaxies are plotted versus spectroscopic redshift in Figure \ref{FIG:PhotozVsSpecz}, and compared to the distribution of photometric redshift errors for the larger sample of 1285 $K$ selected galaxies with both redshifts available.  1$\sigma$ and 2$\sigma$ regions are indicated by shading in the figure.  Three s{\it BzK} galaxies have photometric redshifts that are more than 2$\sigma$ outliers; this outlier fraction is consistent with Gaussian statistics given our sample size.

The final sample of 510 star-forming s{\it BzK}s with redshift $1.2<z\leq3.2$ are binned according to redshift into approximately 1 Gyr intervals in cosmic time:  $1.2<z\leq1.5$, $1.5<z\leq2.0$ and $2.0<z\leq3.2$; details are included in Table \ref{TABLE:RedshiftBinnedSample}.   These sets of galaxies are used in analysis of individual detections and combined non-detections (stacking analyses) to determine their aggregate fluxes in each waveband.  

The redshift binned sets are illustrated in a {\it BzK} diagram in Figure \ref{FIG:BzKDiagram}.  ($B$ - z) and (z - $K$) colors for these galaxies are determined from flux values given in the catalog of \citet{2008ApJ...681.1099B}.  Non-detections in the $B$ band lead to undetermined ($B$ - z) colors for 4 and 13 galaxies in the redshift bins $1.5<z\leq2.0$ and $2.0<z\leq3.2$, respectively.  These galaxies are excluded from Figure \ref{FIG:BzKDiagram}.  

%
\section{MULTIWAVELENGTH ANALYSES AND STACKING ANALYSES}
\label{MethodSection}

We measure the redshift-bin-averaged multiwavelength SEDs of our sample, and use these data to estimate SFRs using published broadband calibrations.  Most of these galaxies are not individually detected in far IR-radio and X-ray wavebands; therefore stacking analysis, using the $K$ band positional priors, is essential. The same analysis method is used for stacking in each IR-radio waveband:  an ordinary average of fluxes of individual detections and the stacking detection yields an aggregate flux estimate for each redshift bin.  As discussed in the Appendix, a weighted average of individual and stacking detections may introduce a bias toward dim sources.   The same set of galaxies are analyzed in all wavebands with the exception of X-ray, in which only sources in the CDF-S are studied.  We do not expect the additional position selection criterion of these sources to introduce any bias.   Furthermore, as discussed in Section \ref{X-RayLuminositySection}, galaxies in the CDF-S will dominate the stacking signal of sources in the full ECDF-S due to the greater X-ray exposure in CDF-S.   

%
%
\subsection{IR-Submillimeter}
\label{IRLuminositySection}
{\bf MIPS 24, 70 $\mu$m.}  Infrared data were obtained from the {\em {\it Spitzer} Space Telescope} Multi Band Imaging Photometer (MIPS) 24 $\mu$m and 70 $\mu$m  images that reach 5$\sigma$ depths of 50 $\mu$Jy and 3 mJy, respectively \citep{2009A&A...496...57M}.  The $K$ band s{\it BzK} positions are compared to the 24 $\mu$m (70 $\mu$m) catalog, with positions less than 2$''$ (4$''$) separation as the criterion for matching to individual detection in each MIPS band, respectively.  Individual detections are removed from the list for stacking, and incorporated into the analysis subsequently.  In 24 $\mu$m data, in redshift bins from $1.2<z\leq1.5$, $1.5<z\leq2.0$ and $2.0<z\leq3.2$, there are 64, 90, and 61 individual detections and 92, 125, and 78 stacked positions, respectively.  Similarly, in 70 $\mu$m data, in these same redshift bins, there are 4, 9, and 1 individual detections and 151, 205, and 136 stacked positions in each of the redshift bins (1, 1, and 2 sources, respectively, could not be analyzed due to their being on the edge of the 70 $\mu$m image).  Stacking is performed on a residual image, after removing the matched sources from the list, as summarized in \citet{2007ApJ...659..305H}.  The stacking algorithm computes an inverse variance weighted average of the flux at each stack position. We use an ordinary average of the individual detections and the stacking detection to yield a single combined estimate for the flux of each set of galaxies.  We show in the Appendix that, although the difference between these two approaches is small, the ordinary average is preferred because the weighted average can introduce a bias to the combined flux estimate. 

{\bf BLAST 250, 350, 500 $\mu$m}.  Submillimeter data at 250, 350, and 500 $\mu$m were obtained from the public archive of the Balloon-borne Large Area Space Telescope (BLAST) survey of the ECDF-S, which reaches 1$\sigma$ depths of 36, 31, and 20 mJy at 250, 350 and 500 $\mu$m, respectively, in an 8.7 deg$^{2}$ wide field and 1$\sigma$ depths of 11, 9, and 6 mJy at 250, 350, and 500 $\mu$m in a 0.8 deg$^{2}$ deep field \citep{2009Natur.458..737D}.  

The redshift binned s{\it BzK}s are stacked in the public BLAST `smooth' data (variance-weighted correlation between the signal maps and the effective point-spread functions).  Each pixel in these data products represents the maximum likelihood flux density (Jy) of an isolated point source centered over the pixel \citep{2008ApJ...681..415T}.  

We use an improved submillimeter stacking and deblending algorithm for stacking in 250, 350, 500 and 870 $\mu$m data that deals effectively with the problem of confusion \citep{2010AJ....139.1592K}.  Confusion severely limits the effectiveness of stacking in deep surveys with limited angular resolution  \citep{1974ApJ...188..279C, 2001AJ....121.1207H}, particularly at far IR-submillimeter wavelengths, and causes a bias in stacking results.  Deblending corrects measured fluxes for confusion from these adjacent sources.  This stacking and deblending algorithm greatly reduces bias in the flux estimate with nearly minimum variance.  For more details, see \citet{2010AJ....139.1592K}.  All galaxies in the MUSYC catalog with $K_{AB}<22$ are used in the deblending calculations.  

We find stacking detections (defined as SNR $\geq$ 3) in the 250 $\mu$m data for the redshift bin $1.5<z\leq2.0$ (stacking detection SNR = 12), and in the 350 $\mu$m data for the redshift bins $1.2<z\leq1.5$ (SNR = 3) and $1.5<z\leq2.0$ (SNR=10), and in the 500 $\mu$m data, for the redshift bin $1.5<z\leq2.0$ (SNR=10).  See Table \ref{Table:FluxDensity} for the stacked flux densities and errors in each waveband.  In the SED fits discussed below, the measured fluxes of formal non-detections and their appropriate error bars are included in the fits.  Combining all of the {\it BzK} galaxies, without regard to redshift binning, yields stacking detections in 250, 350, and 500 $\mu$m data of 3.9$\pm$0.4 mJy, 2.5$\pm$0.3~mJy, and 1.8$\pm$0.2~mJy respectively.  

Fluxes reported from stacking 24 $\mu$m selected {\it BzK} galaxies in the same field, with a different stacking algorithm are larger by about a factor of two than those values presented here \citep{2009arXiv0904.1205M}; in addition to the difference in selection of the present sample (which is $K$ selected and excludes AGN), this discrepancy may also possibly be attributed to lack of deblending in these previous reported results \citep{2010arXiv1003.1731C}.  

{\bf LESS 870 $\mu$m}.  Submillimeter data at 870 $\mu$m were obtained from the Large Apex Bolometer Camera ECDF-S Submillimeter Survey (LESS;  \citealt{2009arXiv0910.2821W}), which reaches a 1$\sigma$ depth of approximately 1.2 mJy beam$^{-1}$.  The LESS catalog contains 126 individually detected submillimeter sources \citep{2009arXiv0910.2821W} and these data have been used previously for stacking analyses of {\it BzK} galaxies \citep{2010ApJ...719..483G}.  

The redshift binned s{\it BzK}s are stacked in the beam-smoothed, flux map  \citep{2009arXiv0910.2821W}; galaxies in the MUSYC $K$ band catalog are used in the deblending calculations.   Individual detections, as identified through 1.4 GHz and/or MIPS 24 $\mu$m counterparts \citep{2011MNRAS.413.2314B}, are excluded from the stacking/deblending analysis and incorporated into the aggregate (stacking + individual detections) flux estimates as discussed above and in the Appendix.
There were 2, 6, and 1 individual 870 $\mu$m detections in the $1.2<z\leq1.5$, $1.5<z\leq2.0$ and $2.0<z\leq3.2$ bins.  These individual detections contributed 46\%, 29\%, 9\% to the aggregate (individual + stacking) detection, respectively.  Stacked flux estimates are $<$282 (3$\sigma$), 509$\pm$80, and 533 $\pm$100 $\mu$Jy in the $1.2<z\leq1.5$, $1.5<z\leq2.0$ and $2.0<z\leq3.2$ bins.  These stacked flux estimates are combined with individual fluxes into the aggregate values indicated in Table \ref{Table:FluxDensity}.

{\bf IR Luminosity Estimation.}  In order to estimate L$_{IR}$, we fit observed IR-radio photometry to template libraries from \citet{2001ApJ...556..562C}.  We explore several approaches; in each case, a different region of the IR-radio spectrum is chosen for template fitting.  Fits are performed on each redshift bin-averaged spectrum.  The template rest frame luminosity is converted to an observed frame flux distribution at the median redshift of the redshift bin.  Then the observed frame model flux distribution is convolved with each photometric bandpass transmission function to generate predicted model photometry.  The predicted photometry are combined with observed photometric fluxes and errors to generate a $\chi^2$ statistic for each fit.    Each template in the library is fit in this way, with the smallest $\chi^2$ fit chosen as the best fit template.  For fits that include observations at multiple wavelengths, an overall normalization, $A$, set to its analytical best fit value via $\partial \chi^2 / \partial A = 0$, is factored into the best fit spectrum.

We explore several additional approaches to estimating L$_{IR}$ from the data using CE01 template fits:  for each redshift bin, we fit (1) the 24 $\mu$m and longer wavelength data, (2) the long wavelength IR and radio data excluding 24 $\mu$m, (3) 24 $\mu$m only data. Optical/NIR data are excluded from the fits because CE01 templates are considered to be incomplete for $\lambda < 1 \mu$m.  For the 24 $\mu$m only fits, there is no free normalization factor, and the CE01 template luminosity is used directly to estimate L$_{IR}$.  For these single band fits, the variation of $\chi^2$ with template index is used as the basis for determining confidence intervals; 68\% error bars  are found for the L$_{IR}$ estimate, and these errors are propagated into an SFR uncertainty.  These fits are illustrated in Table \ref{TABLE:IRFITSUMMARY} and Figure \ref{FIG:CE01TemplateFits} and are discussed in Section \ref{ResultsSection}.

Uncertainties in L$_{IR}$ are determined from 68\% confidence intervals determined from  variations in $\chi^2$ with normalization.  We considered separately the effects of error in the redshift.  We approximate the error of the median redshift by computing the error of the mean of individual redshifts.  This error of the mean diminishes according $1/\sqrt N$, where $N$ is the number of objects in each redshift bin.  We computed L$_{IR}$ for our samples at the $\pm 1\sigma$ values of the mean redshift and found the results to be the same as the actual L$_{IR}$ to within the normalization error.  Consequently, the error in the mean redshift does not contribute significantly to the overall L$_{IR}$ uncertainty.  Because the median redshift is more robust to the presence of outliers, we consider the error of the mean to be an upper bound to the error of the median.  We also address the effect of redshift and individual galaxy SED errors via simulations, as discussed in Section \ref{RedshiftSmearingSection}.  On this basis, we ignore error of the median redshift in subsequent calculations. 

%
We address the question of whether L$_{IR}$ determined from the average flux SED is indicative of the true average of individual galaxy luminosities in two ways:  (1) using observations of the (bright) subset of 24 $\mu$m detected sources, and (2) in simulations, for the entire sample including individually non-detected sources.

The redshift bin average L$_{IR}$ for individually detected 24 $\mu$m sources is computed based on individual fits to CE01 templates, using only the 24 $\mu$m band photometry.   Average L$_{IR}$ is also computed for this 24 $\mu$m bright subset using the same procedure as used in the larger, sample of {\it BzK}s:  flux values are combined to form an unweighted average SED.  This average SED is then fit to CE01 templates to determine L$_{IR}$.  The two  L$_{IR}$ estimates agree to within 10-20\% in our redshift bins.   In Section \ref{RedshiftSmearingSection}, we generalize these results to our entire sample using simulations.

We use the calibration of \citet{1998ApJ...498..541K} to convert L$_{IR}$ to estimated SFR.  This calibration is based on the starburst synthesis models of \citet{1995ApJS...96....9L}, and it assumes a continuous burst of age 10-100 Myr, solar abundances, Salpeter IMF, and bolometric luminosity arising from dust reradiation.  This calibration relates the L$_{IR}$ integrated from 8-1000 $\mu$m to SFR according to
\begin{equation}
SFR_{IR}  (M_\odot~yr^{-1}) = 4.5 \times 10^{-44} L_{IR} (\text{erg~s}^{-1})
\label{EQ:KennicutIRSFR}
\end{equation}
The uncertainty in this relation arises from uncertainties in estimation of $L_{IR}$ resulting from extrapolation of observed fluxes to total, integrated $L_{IR}$ , confounding sources of IR emission that are not associated with star formation, and the use of a fixed continuous burst model; the combined errors in the SFR are attributed as being a factor of $\sim 2-3$.  This systematic uncertainty in luminosity to SFR conversion dominates the overall error budget in SFR estimates; the impact of this substantial systematic uncertainty on other SFR estimates, many of which depend indirectly on the L$_{IR}$-SFR relationship, is discussed in Section \ref{DiscussionSection}.  SFR estimates for our redshift binned s{\it BzK}s are discussed in Section \ref{IRSFREstimates}.   Uncertainties in these SFR estimates include only errors in L$_{IR}$ (arising from photometry, and template normalization). 

\subsection{UV-Optical-NIR}
\label{UVSFRSection}

MUSYC 5$\sigma$ imaging depths include U = 26.5, B=26.9, V=26.6, R=26.3, I=24.8, z$'$=24.0, J =23.1, H=22.4, K=22.4 as well as 18 medium band photometry in the range [4270,8560] \Angstrom \citep{2006ApJS..162....1G,2009ApJS..183..295T,2010ApJS..189..270C}.  Galaxies are individually measured in MUSYC $UBVRIz'JHK$ bandpasses via aperture photometry, and additionally their fluxes in each redshift bin are combined in an unweighted average to yield a single averaged  SED for each redshift range.  These average fluxes for redshift binned s{\it BzK}s in each UV-radio waveband are indicated in Table \ref{Table:FluxDensity}.  IR data in the ECDF-S are available in IRAC bands at 3.6, 4.5, 5.8 and 8.0 $\mu$m (SIMPLE; \citealt{2011ApJ...727....1D}).  As discussed in Section \ref{DataSection}, IRAC data were used in photometric redshift determination; however, these data were not used in the fits to determine $L_{IR}$.

To determine UV continuum luminosity before dust correction, $L_\nu^{Uncorr}$, we use the available optical-NIR photometry.  We estimate the rest frame 1600 \Angstrom flux density, $f_\nu^{Uncorr}$, via interpolation of the two bracketing broadband fluxes.  The specific luminosity at 1500 \Angstrom, $L_\nu^{Uncorr}$ in units of erg~s$^{-1}$ Hz$^{-1}$, at the redshift, $z$, is then found from the flux density, $f_\nu^{Uncorr}$ in units of $\mu$Jy and the luminosity distance, $D_L$, according to
\begin{equation}
L_\nu^{Uncorr} = 1 \times 10^{-29} f_{\nu}^{Uncorr} \frac{4\pi D_L^2}{(1+z)}.
\end{equation}

To convert luminosity to SFR, we use the calibration of \citet{1998ARA&A..36..189K}, which corresponds to the calibration of \citet{1998ApJ...498..106M} converted to Salpeter IMF and 0.1 -- 100 M$_\odot$ mass limits.  
\begin{equation}
SFR_{UV}(M_\odot~yr^{-1}) = 1.4 \times 10^{-28}  L_\nu(\text{erg~s}^{-1} Hz^{-1})
\label{EQ:UVSFR}
\end{equation}
This calibration assumes continuous star formation over timescales of 10$^8$ years or longer and solar metallicity.   \citet{1998ARA&A..36..189K} discusses sources of systematic uncertainty in the various  published $L_\nu$ -- SFR calibrations as arising from the use of different stellar libraries and assumptions about the star formation time scales; the published calibrations differ by about a factor of 2.  We use Equation \ref{EQ:UVSFR} to compute SFR from uncorrected luminosities, $L_\nu^{Uncorr}$, to estimate the contribution to SFR that is unobscured by dust.  We also apply this expression to dust-corrected luminosities to determine dust-corrected UV SFR, SFR$^{Corr}_{UV}$, as discussed below. 

To obtain dust-corrected UV SFRs, we use the method of IRX-$\beta$ \citep{1999ApJ...521...64M}.  This method has been used in high redshift LBGs, (e.g. \citealt{1999ApJ...521...64M,2000ApJ...544..218A,2010ApJ...712.1070R}), {\it BzK} galaxies \citep{2007ApJ...670..156D}, and galaxies at lower redshift (e.g. \citealt{2010MNRAS.409L...1B,2010ApJ...715..572H,2010A&A...514A...4T}).  

We fit the rest frame UV $f_\lambda$ spectrum to a power law,  $f_\lambda(\lambda) = A \lambda^\beta$, using a range of trial $\beta$ values: -2.5$<\beta<$1.0 in steps $\Delta \beta$ = 0.01. Due to the availability of 18 medium band photometry, these fits typically had 12-14 sampled points in the spectrum.  The wavelength range of the fits,  [1268, 2580]  \Angstrom in the rest frame, is chosen to be the same as that used in \citet{1994ApJ...429..582C}, which \citet{1999ApJ...521...64M} also adopted.  These values are redshifted into the observed frame, and photometry data falling within this range are used for the fits.  For each trial value of $\beta$, predicted flux values are computed at each relevant, observed wavelength by integrating over the filter bandpass transmission function, $T(\lambda)$, and intergalactic medium transmission function, $M(\lambda)$, from \citet{1995ApJ...441...18M}.  The integral is expressed in terms of the number of photons detected, hence an extra factor of $\lambda$ is included in the integrand, as illustrated below in Equation \ref{eq:FluxIntegral}.  The integral is normalized to units of $\mu$Jy by dividing by the corresponding integral of a reference spectrum that is flat in  $f_\nu$  ($f_{\nu}^{ref}$=1 $\mu$Jy), which is converted to a photon number spectrum.  This approach leads to the expression for predicted flux density, $f_{\nu_{i}}^{pred}$, for each broadband filter, $i$
\begin{equation}
f_{\nu_i}^{pred}(\mu Jy) = A \frac{\int{\lambda^\beta T_i(\lambda) M(\lambda) \lambda d\lambda}}{\int{f_{\lambda}^{ref}(\lambda)T_i(\lambda) \lambda d\lambda}}
\label{eq:FluxIntegral}
\end{equation}
We compute a $\chi^2$ for each fit, and we optimize the normalization parameter, $A$, by selecting the value for which 
$\partial \chi^2/\partial A = 0$.  Finally the complete, normalized predicted flux in $\mu$Jy is computed from Equation \ref{eq:FluxIntegral}.  The above fit procedure is repeated for each trial $\beta$ value, and the fit with the smallest $\chi^2$ is chosen to represent the data.  

The resulting power law index, $\beta$, is then used to compute the UV extinction from the empirical relation of \citet{1999ApJ...521...64M}
\begin{equation}
A_{1600} = 4.43 + 1.99 \beta,
\label{EQ:MeurerEq11}
\end{equation}
which is found to have 0.55 mag dispersion about their fit in A$_{1600}$ and a standard error in the fit zero point of 0.08 mag; see Equation (11) and Figure 1 from \citet{1999ApJ...521...64M}.  The UV extinction is then used to correct the measured UV flux according to
\begin{equation}
F_{UV}^{Corr} = 10^{0.4 A_{1600}} F_{UV}^{Uncorr}
\end{equation}
Finally the corrected UV flux is used to estimate the SFR using Equation \ref{EQ:UVSFR}.   The above procedure is executed for each s{\it BzK} galaxy individually as well as for the unweighted average spectrum of all s{\it BzK}s within a redshift bin, whereby the bin median redshift is used to compute the luminosity distance.  
 
Uncertainties to the corrected UV luminosity arise from observed flux uncertainty, error in the dust correction factor, and photometric redshift error (for individual galaxies, we adopt the value of $\delta z/(1+z)$ = 0.009; for the average spectra, this contribution is negligible, as discussed in Section \ref{IRLuminositySection} and via simulations in Section \ref{RedshiftSmearingSection}).  These uncertainties are combined using standard error analysis.  We do not include the systematic uncertainty associated with the luminosity$-$SFR calibration (discussed above, about a factor of 2); rather we consider this uncertainty with similar systematics from other waveband estimators separately in Section \ref{DiscussionSection}.  The results of these computations are shown in Table \ref{Table:UVSFR} and discussed in Section \ref{DiscussionSection}.

\subsection{Radio Luminosity and SFR Estimation}
\label{RadioLuminositySection}
{\bf Radio}.  Radio data at 610 MHz were obtained from the Giant Metrewave Radio Telescope (GMRT) survey of the ECDF-S, which reaches a typical depth of 40 $\mu$Jy~beam$^{-1}$ \citep{2010MNRAS.402..245I}. 1.4 GHz data were obtained from the Very Large Array (VLA) survey, which covers the ECDF-S to a typical depth of 8 $\mu$Jy~beam$^{-1}$ and includes 464 cataloged sources \citep{2008ApJS..179..114M}.  

Flux estimates are found from weighted average image stacking (excluding individual detections, which are included after stacking) as well as median image stacking (of all sources) of the VLA and GMRT data.  Median stacking is commonly used to reduce the influence of radio-loud AGN.   We adopt the weighted average method to be consistent with the analysis in other wavebands, and we adopt median stacking for comparison.
In the radio regime, where the spatial resolution is relatively high, making images allows us to conserve flux density that would otherwise be lost due to smearing by astrometric uncertainties and finite bandwidth (chromatic aberration) at the cost of larger flux density uncertainties \citep{2007ApJ...660L..77I}.  Radio fluxes, luminosities and corresponding SFRs are illustrated in Table \ref{TABLE:RADIOSFR}.  

Our data include flux measurements, S$_\nu$, at 1.4 GHz and  610 MHz for each of the redshift binned SEDs.  The radio spectral index, S$_\nu \propto \nu^\alpha$, (typical $\alpha \sim$ -0.8 for { galaxies, e.g. \citealt{1992ARA&A..30..575C}) is estimated for each of our three redshift bins to be -0.74$\pm$0.2, -1.20$\pm$0.2, 0.06$\pm$0.4 respectively.  In the higher redshift bins, these computed indices deviate significantly from the $\alpha$ = -0.8 for synchrotron emission (e.g. anomalously high 610 MHz flux estimate in the $1.5 < z \leq 2.0$ bin). Similarly high fluxes were also reported for galaxies in the range $0<z<2$ in \citet{2011MNRAS.410.1155B}, and interpreted as resulting from AGN contamination at high redshift.  Therefore in keeping with other reported literature, we adopt the $\alpha$ = -0.8 value in computing luminosities and SFRs.

In estimating radio luminosities, we use the median redshift, $z$, and the corresponding luminosity distance, $D_L$, in Mpc to compute the aggregate rest frame 1.4 GHz luminosity, $L_{\nu,\text{1.4 GHz}}$ in units of W~Hz$^{-1}$, from the observed frame 1.4 GHz flux, $S_\nu$, in units of $\mu$Jy according to
\begin{equation}
L_{\nu,\text{1.4 GHz}} = 9.523 \times 10^{12} \frac{4\pi D_L^2}{(1+z)^{1+\alpha}} S_{\nu, \text{1.4 GHz}}
\end{equation} 

To estimate SFR from $L_{\nu,\text{1.4 GHz}}$, we use the model of \citet{1992ARA&A..30..575C} as implemented in \citet{2000ApJ...544..641H} and \citet{2009MNRAS.394....3D}.  Following the implementation in \citet{2000ApJ...544..641H}, SFR in units of M$_\odot$ yr$^{-1}$ is a function of frequency in units of GHz, $L_{\nu,\text{1.4 GHz}}$ in units of W Hz$^{-1}$, scaled by a factor $Q$ and is given by
\begin{equation}
SFR^{Condon}_{1.4 GHz} = Q \frac{L_{\nu,\text{1.4 GHz}}}{5.3 \times 10^{21} \nu^\alpha + 5.5\times10^{20} \nu^{-0.1}}
\end{equation}
We use the value $Q=5.5$ to scale the SFR(M$>$5 M$_\odot$) calculated in  \citet{1992ARA&A..30..575C} to SFR(0.1-100 M$_\odot$) used here; this scaling factor depends on the assumed (Salpeter) IMF used here and by \citet{2000ApJ...544..641H}.

For comparison, we also estimate SFR from 1.4 GHz flux using the calibration of \citet{2003ApJ...586..794B}.  This calibration  is based on the IR-radio correlation; it assumes that nonthermal radio emission directly tracks the SFR, and is chosen so the radio SFR matches the IR SFR for $L\geq L^{*}$ galaxies.  The SFR calibration
\begin{equation}
SFR^{Bell}_{1.4 GHz} (M_\odot~yr^{-1}) = 5.52 \times 10^{-22} L_{\nu,\text{1.4 GHz}}
\end{equation}
is adopted here.  A similar calibration is found in \citet{2001ApJ...554..803Y}.  SFR$^{Condon}_{1.4 GHz}$ exceeds SFR$^{Bell}_{1.4 GHz}$ by a factor of two; the calibration of \citet{1992ARA&A..30..575C} explicitly models the thermal and nonthermal emission mechanisms, whereas the calibration of \citet{2003ApJ...586..794B} relies upon the IR-Radio correlation.  Thus we expect agreement between SFR$^{Bell}_{1.4 GHz}$ and IR-based SFR estimates, if the IR-radio correlation continues to hold at high redshift, as has indeed been suggested in the literature  \citep{2010ApJ...714L.190S,2010A&A...518L..31I}.

Uncertainties to the radio luminosities are computed by incorporating uncertainties from redshift and flux measurement; these uncertainties are propagated into the SFR uncertainties.  When only flux measurement uncertainties are included in the error budget, uncertainties in SFR$^{Condon}_{1.4 GHz}$ agree to within 30\% of published values \citep{2009MNRAS.394....3D}.  In the calibration of \citet{2003ApJ...586..794B}, scatter in the IR-radio correlation contributes a factor of 1.8 (dispersion of 0.26 dex for individual galaxies) to the uncertainty and dominates the total error budget; this additional systematic uncertainty arising from the $L_{IR}$--SFR calibration is discussed in Section \ref{DiscussionSection}

%
\subsection{X-Ray SFR Estimation}
\label{X-RayLuminositySection}
X-ray exposure in the ECDF-S consists of 4 Ms in the central $\approx$ $16'\times16'$ CDF-S, reaching approximate sensitivities of 1$\times$10$^{-17}$ and 7$\times$10$^{-17}$ erg cm$^{-2}$ s$^{-1}$ in the 0.5-2.0 and 2.0-8 keV bands, respectively, and giving this field the deepest X-ray coverage to date \citep{2011ApJS..195...10X}.  
These data are augmented with four flanking 250 ks exposures that complete the $\approx 30' \times 30'$ ECDF-S field, and reach sensitivity limits of 1.7$\times$10$^{-16}$ and 3.9 $\times$10$^{-16}$ erg cm$^{-2}$ s$^{-1}$ in the 0.5-2.0 and 2.0-8.0 keV bands, respectively \citep{2005ApJS..161...21L,2006AJ....131.2373V}.   

X-ray stacking analysis was performed in the 4 Ms CDF-S; due to the ratio of exposure times in the CDF-S versus ECDF-S,  the deeper CDF-S data will dominate any stacking signal.  The X-ray stacking algorithm is discussed in \citet{2011Natur.474..356T}; a position dependent aperture correction was used to account for the varying {\it Chandra} PSF with off-axis angle, and to minimize this correction, only sources within 10$'$ of the aim point were stacked.  Sources that have an X-ray detection closer than 15$''$ to the stacking position are removed to provide a better estimation of the background.  This procedure leaves 19, 29, and19 source positions in redshift bins from $1.2<z\leq1.5$, $1.5<z\leq2.0$ and $2.0<z\leq3.2$ respectively that are stacked in {\it {\it Chandra}} soft band and hard band data.  Stacked fluxes are combined with one individually detected source each in the $1.2<z\leq1.5$ and $1.5<z\leq2.0$ bins, according to the procedure in the Appendix.  Including these individual detections increased the soft band stacking flux estimate by 33\% in the $1.2<z\leq1.5$ bin and had negligible effect on the $1.5<z\leq2.0$ bin.

Counts to flux, and flux to luminosity conversions are done assuming a spectrum with photon index $\Gamma$=1.2 and cutoff energy, $E_c$=20 keV.  The rest frame 2-10 keV luminosities are used to estimate SFR.  
As shown by \citet{2002A&A...382..843P}, the X-ray spectrum of star-forming galaxies that do not have an AGN is dominated by HMXBs, which are best described by $\Gamma=1.2$ and a cutoff energy of 20 keV.  Many star-forming galaxies also present a thermal component, which is typically softer in X-rays with $kT\sim0.7$ keV \citep{1989ARA&A..27...87F}.  The spectrum of the resulting combination is something softer than a pure  $\Gamma=1.2$, but not quite $\Gamma=2$.  In studying LBGs, \citet{2002ApJ...576..625N} assumed an intrinsic spectrum of $\Gamma=2.0$, more typical of local Seyfert galaxies and soft X-ray selected quasars.  To estimate the effect of different assumptions of photon spectrum index, we also computed counts to flux and luminosity conversions using $\Gamma$ = 1.9.  The differences of conversions from counts to flux are $\sim$ 11\%.  Similarly, the differences in conversion from observed frame soft band to rest frame hard band are $\sim$ 16\% between these two assumptions of spectral index.  Thus in the soft band, the uncertainties due to an assumed spectral shape are $\sim$ 20\%.  X-ray fluxes, luminosities and SFRs are tabulated in Table \ref{TABLE:XRAYSFR} and discussed in Section \ref{XraySFRSection}.

The X-ray - SFR calibration of \citet{2003A&A...399...39R} is widely used, and is based upon the X-ray - IR luminosity correlation  observed in galaxies with L$_{2-10 keV} \lesssim 10^{41}$ erg s$^{-1}$.  The SFR in units of $M_\odot$ yr$^{-1}$ is related to the 2-10 keV luminosity, L$_{2-10 keV}$ in units of erg s$^{-1}$, according to
\begin{equation}
SFR^{Ranalli}_{2-10 keV}=2.0\times10^{-40} L_{2-10 keV}.
\label{Eq:Ranalli_No2}
\end{equation}
This calibration implicitly assumes the $L_{IR}$--SFR calibration of \citet{1998ARA&A..36..189K} and a Salpeter IMF (0.1-100 M$_\odot$) consistent with other calibrations mentioned in this paper.\footnote{Note that the radio SFR calibration cited in \citet{2003A&A...399...39R} refers to M$>$ 5 M$_\odot$ mass range.  For a Salpeter IMF, the resulting X-ray--radio derived SFRs differ by a factor of 5.5 from the 0.1-100 M$_\odot$ range used here.}  However, the \citet{2003A&A...399...39R} sample includes few star-forming galaxies in the ULIRG regime, where the $L_X$ -- SFR correlation is observed to drop \citep{LehmerPrivComm2010}.

The uncertainty to SFR$^{Ranalli}_{2-10 keV}$ is computed by adding in quadrature uncertainties in X-ray luminosity and the 0.09 dex error of the slope in the X-ray$-$IR-luminosity correlation (see  \citealt{2003A&A...399...39R} Equation 10).  Luminosity uncertainties are computed by propagating the flux estimate errors; redshift errors of the average spectrum can be neglected, as discussed in Section \ref{IRLuminositySection} and shown in simulations discussed in Section \ref{RedshiftSmearingSection}.

Subsequent studies have related instantaneous SFR specifically to luminosity from short-lived (e.g. \citealt{2003MNRAS.339..793G,2004ApJ...602..231C,2004A&A...419..849P}) while slowly evolving, LMXBs are linked to stellar mass, i.e. integrated SFR \citep{2004ApJ...602..231C}.  The X-ray SFR calibration of \citet{2004A&A...419..849P} is based upon the luminosity of HMXBs, and it relates SFR in units of $M_\odot$ yr$^{-1}$  to the 2-10 keV HMXB luminosity, L$_{2-10 keV}^{HMXB}$ in units of erg s$^{-1}$, according to
\begin{equation}
SFR_{2-10 keV}^{Persic} = 10^{-39} L_{2-10 keV}^{HMXB}.
\label{Eq:Persic}
\end{equation}
The fraction, $f$, of HMXB X-ray luminosity to total X-ray luminosity has been estimated as $f\sim 0.2$ (with substantial scatter due to low statistics) for nearby star-forming galaxies \citep{2004A&A...419..849P}.  For high redshift ($z>1$)  galaxies, in the absence of definitive estimates from X-ray spectroscopy, the value $f=1$ has been used on the assumption that LMXBs (or other sources of emission) contribute a negligible fraction to the total X-ray luminosity at $z\sim2$ \citep{2004A&A...419..849P,2007A&A...463..481P}.

Assuming $f=0.2$ for nearby star-forming galaxies leads to L$_{2-10 keV}^{HMXB}$ = 0.2 $L^{Total}_{2-10 keV}$, which brings the calibration of \citet{2004A&A...419..849P} into equivalence with the calibration of \citet{2003A&A...399...39R}.  
Our data for high redshift galaxies do not directly constrain the HMXB luminosity fraction; with the assumption $f=1$, $L_{2-10 keV}^{HMXB}$ = $L^{Total}_{2-10 keV}$, and the SFRs estimated from  \citet{2004A&A...419..849P} exceed those of  \citet{2003A&A...399...39R} by a factor of 5.  In computing SFR, we adopt $f=1$ for the X-ray calibration of \citet{2004A&A...419..849P} for our sample of s{\it BzK}s, and we regard the resulting SFRs as upper limits. 

The relative contribution of LMXBs to the total X-ray luminosity is believed to decline above $z\sim1$ \citep{2001ApJ...559L..97G}, and to be subdominant in high SFR (e.g. $>100$ M$_\odot$ yr$^{-1}$) galaxies.  A bilinear relation between X-ray luminosity and both SFR and stellar mass, $M_\star$, has been proposed, (e.g. \citealt{2004ApJ...602..231C}).  The X-ray SFR calibration of \citet{2010ApJ...724..559L} is derived from such a relationship
\begin{equation}
L_{2-10 keV}^{Lehmer} = \alpha M_\star + \beta SFR.
\label{Eq:Lehmer}
\end{equation}
In analysis of LIRGs/ULIRGs extending to L$_{2-10 keV}$ $\sim 10^{41.5}$ erg s$^{-1}$, \citet{2010ApJ...724..559L} report $\alpha = (9.05\pm 0.37) \times 10^{28}$ erg s$^{-1}$ M$_\odot^{-1}$ and $\beta = (1.62\pm 0.22) \times 10^{39}$ erg s$^{-1}$ (M$_\odot$ yr$^{-1}$)$^{-1}$.  In the absence of the M$_\star$ term in Equation \ref{Eq:Lehmer}, and with the assumption $f=1$, as discussed above, this calibration becomes consistent to within errors of the calibration of \citet{2004A&A...419..849P}, after accounting for the differences in IMF assumed by these authors.

In order to compute SFR$^{Lehmer}_{2-10 keV}$, we estimate stellar masses for our redshift binned s{\it BzK} samples.  We use the empirical correlation between the observed frame $K$ band magnitude and the stellar mass for s{\it BzK}s at $z>1.4$, determined from SED fits, that is presented in \citet{2004ApJ...617..746D}.
\begin{equation}
\log(M_\star / 10^{11} M_\odot) = -0.4(K^{tot} - K^{11})
\label{Eq:Daddimass}
\end{equation}
where $K^{11} = 21.4$ is the $K$ band magnitude corresponding on average to a stellar mass of 10$^{11}$ M$_\odot$.   \citet{2004ApJ...617..746D} report uncertainties of $\sim$40\% with this relation.

Uncertainties in SFR$^{Lehmer}_{2-10 keV}$ are calculated by propagating the uncertainties in the luminosity and stellar mass, along with the reported uncertainties in the parameters $\alpha$ and $\beta$ indicated above.  This calibration is based implicitly on the $L_{IR}$--SFR calibration of \citet{2005ApJ...625...23B} which yields lower SFRs by $\approx$13\% compared to the corresponding calibration of \citet{1998ApJ...498..541K}; however, we neglect this small calibration difference, and discuss systematic uncertainties in comparison with other SFR indicators in Section \ref{DiscussionSection}.

%
\subsection{SIMULATIONS}
\label{RedshiftSmearingSection}
We investigate the effects of redshift bin averaging, photometric redshift errors and dispersion of individual galaxy SEDs on our stacked L$_{IR}$ estimates with simulations.  Averaging the photometric flux density from galaxies at slightly different redshifts within a redshift bin introduces redshift smearing.  In order to study this effect, we simulate a set of identical CE01 template spectral models.  These spectra are shifted to the identical redshifts of galaxies in our $1.5<z\leq2.0$ redshift bin, and then averaged together, analogous to the actual stacking procedure.  The redshift bin averaged spectrum is nearly identical to the template spectrum except for moderate smearing of the emission peaks that contributed a negligible amount to the integral; consequently, the quantity of interest, the integrated IR luminosity, is robust against redshift smearing over our bin widths.  

Of greater concern is the effect on L$_{IR}$ due to photometric redshift errors and dispersion due to individual galaxy SEDs.  To quantify the contribution of these errors to the estimated L$_{IR}$ for each redshift bin, we simulate sets of galaxies with SEDs chosen at random from CE01 templates and distributed in redshift to simulate the observed source distribution.  The photometric redshift error distribution is determined from comparison of spectroscopic and photometric redshifts for the subset of sources with both estimates available and is shown in Figure \ref{FIG:PhotozVsSpecz}.  This distribution is well fit by a Gaussian with mean = -0.02 (i.e. bias) and $\sigma$=0.09 (i.e. scatter).  The bias is first removed from the simulated object redshifts, and then artificial redshift errors drawn from this biased, Gaussian distribution are added in each repeated trial of the simulation.  The resulting spectra are averaged, and this averaged spectrum is integrated to determine L$_{IR}$.  
An example of these spectra from the $1.5<z\leq2.0$ redshift bin simulation is illustrated in Figure \ref{FIG:ZBINSIMULATION}.  From the figure, it is apparent that the averaged spectrum has a slightly higher flux than the single object spectrum near the emission peak and therefore will overestimate L$_{IR}$.  For each repeated trial, the fractional error in the L$_{IR}$ estimate is determined by comparing the bin averaged L$_{IR}$ to the true redshift bin averaged luminosity.  The frequency distribution of L$_{IR}$ fractional errors is determined directly from 10$^4$ repeated trials for each redshift bin.  The fractional error distributions for redshift bins $1.2<z\leq1.5$, $1.5<z\leq2.0$ and $2.0<z\leq3.2$ are each Gaussian with mean = 12\%, 14\%, 2\% (bias) and $\sigma$ = 7\%, 6\%, 6\% (scatter) respectively.  The fractional error distribution for the $1.5<z\leq2.0$ bin is illustrated in Figure \ref{FIG:ZBINSIMULATION}.  Our reported values in Table \ref{TABLE:IRFITSUMMARY} are bias subtracted, and have the scatter added in quadrature with other sources of errors.

%
Finally, in simulations we address the issue of whether luminosities computed from the average flux SED taken to be at the median redshift of each bin may accurately reflect the true average luminosity of our sample of individual galaxies.  To test this method, we distribute a set of galaxies, with CE01 templates chosen at random, distributed in redshift according to the actual source population, compute the L$_{IR}$ of each galaxy individually, and average them to determine true average L$_{IR}$.  Then for each galaxy, we compute observed frame photometric fluxes in each FIR-radio waveband, and subsequently compute the sample average observed flux SED.  We compute the L$_{IR}$ of the average flux SED, assumed to be at the median redshift using the methods of Section \ref{IRLuminositySection}.  We compare this L$_{IR}$ estimate with the true average L$_{IR}$, and determine the distribution of errors with 10$^3$ Monte Carlo realizations.

The fractional error distributions for redshift bins $1.2<z\leq1.5$, $1.5<z\leq2.0$ and $2.0<z\leq3.2$ are each Gaussian with mean = 0.02, 0.09, and 0.02 respectively and $\sigma$ = 0.03 in each case.  Thus the average flux spectrum approximation introduces only a small redshift dependent bias (which may be removed by using an rms effective redshift for each bin) and a scatter of $\sim$3\% to our L$_{IR}$ estimates.  These errors are small especially compared to systematics; therefore our samples of redshift binned galaxies are well represented by an average flux SED at the median redshift.

\section{RESULTS}
\label{ResultsSection}
%
\subsection{IR SFR Estimates}
\label{IRSFREstimates}
SFR$_{IR}$ values that are obtained from CE01 template fits are illustrated in Figure \ref{FIG:CE01TemplateFits}, and tabulated in Table \ref{TABLE:IRFITSUMMARY}.    All of our redshift bins contain significant submillimeter detections, which help to constrain the dust emission peaks.  We adopt CE01 template fits in the range $\lambda \geq 24 \mu$m for our preferred L$_{IR}$ values.  Though not formally the best $\chi^2$, they are comparable to the best fits and including the 24 $\mu$m photometry makes maximum use of the available data.  L$_{IR}$ values obtained from Table \ref{TABLE:IRFITSUMMARY} are consistent with results of \citet{2005ApJ...631L..13D}, who select {\it BzK}s to the same depth in the $K$ band as presented here, and use MIPS 24 $\mu$m photometry to estimate L$_{IR} \sim 1.7 \times 10^{12}$ erg s$^{-1}$ for {\it BzK}s in the range $1.4<z<2.5$. 

%
\subsection{UV SFR Estimates}
\label{UVSFREstimates}
Table~\ref{Table:UVSFR} illustrates the unweighted averages of estimates of UV SFRs from analysis of individual s{\it BzK} galaxies in each redshift bin.  Averages excluded galaxies with poor fits to the spectral index, $\beta$, identified by large $\chi^2$ ($\chi_\nu^2 > 2$) or best fit $\beta$ values that were pinned at the extreme of the allowed parameter range.  In our three redshift bins $1.2<z\leq1.5$, $1.5<z\leq2.0$, and $2.0<z\leq 3.3$, these poor fit criteria excluded 41, 82, and 48 galaxies, respectively, from the averages.  Average SFR$_{UV}^{Corr}$ is in the range 12--285 M$_\odot$ yr$^{-1}$, increasing with redshift.  SFR in the highest redshift bin is affected by outliers; the median SFR$_{UV}^{Corr}$ for individual galaxies are 10, 33, and 106 M$_\odot$ yr$^{-1}$ in each redshift bin, respectively.  However, in keeping with the literature, we adopt the average of individual fits as our preferred indicator of UV SFRs for our sample.

We also compute the SFRs from a single unweighted average spectrum of galaxies within each redshift bin.  These estimates are systematically lower than the averages of individual galaxies in each bin because the best fit UV continuum slopes to the average spectra indicate a lower dust correction than the average of individual fits.  For redshift bins $1.2 < z \leq 1.5$, $1.5 < z \leq 2.0$, and $2.0 < z \leq 3.3$, the fits to average spectra had reduced $\chi^2$ values of 0.3, 1.8, and 3.8 respectively.   We interpret these values to mean acceptable fits for the lower two redshift bins.  

There are eleven galaxies in the highest redshift bin with SFR $>$1000 M$_\odot$ yr$^{-1}$.  Checking the positions of these galaxies against the published LESS catalog \citep{2009arXiv0910.2821W} indicates that they are not submillimeter sources; separations between these galaxies and their nearest neighbor in the submillimeter catalog are all greater than 50$''$.  Five of these sources are detected in 24 $\mu$m waveband, and their inferred luminosities and SFRs (from CE01 fits) are also high (SFR$_{24 \mu m} >$600 M$_\odot$ yr$^{-1}$); however, as discussed below, SFR$_{24 \mu m}$ is known to be overestimated in this redshift and luminosity range.  One of these sources is detected in radio, with $L_{1.4 GHz} = 3 \times 10^{24}$ W Hz$^{-1}$ (SFR$_{1.4 GHz}^{Bell}$ = 1650 M$_\odot$ yr$^{-1}$), and therefore may be an example of previously reported Optically Faint Radio Galaxies (OFRGs; \citealt{2004ApJ...614..671C,2009MNRAS.399..121C}).  

IRAC colors are available for three of these eleven galaxies, and none of them appear in the AGN selection region of \citet{2005ApJ...631..163S} in IRAC color-color space. None of these sources are individually detected in X rays, although three of them are within the CDF-S (between 5-10$'$ from center).  AGN contamination cannot be ruled out; however, we would expect obscuration of the AGN in rest-frame UV and optical wavebands to mean that star formation would dominate the emission (as opposed to the case in X-ray wavebands, where obscured AGN are a dominant confounding factor).  These outliers may suggest either different dust physical properties or geometry in these galaxies.

In comparison with other literature works, \citet{2004ApJ...617..746D} determine SFRs for a $K_{Vega} < 20$ sample of 24 s{\it BzK} galaxies in the GOODS-S field at $z>1.4$ using SED fitting and dust correction using the method of  \citet{1999ApJ...521...64M}, and find dust-corrected SFR in the range 100--600 M$_\odot$ yr$^{-1}$.  In a spectroscopically selected sample of {\it BzK}s, \citet{2010ApJ...718..112Y} find SFRs to vary widely, over three orders of magnitude.

\subsection{Radio SFR Estimates}
\label{RadioSFRSection}
Radio fluxes, luminosities and associated SFRs are reported in Table \ref{TABLE:RADIOSFR} and radio SFRs are compared to calibrations in other wavebands in Table \ref{Table:SFR}.  It has been reported previously that SFR$^{Condon}_{1.4 GHz}$ exceeds SFR$^{Bell}_{1.4 GHz}$ by approximately a factor of two \citep{2003ApJ...586..794B}.   Discrepancies between the radio SFR calibrations of \citet{2003ApJ...586..794B} and \citet{1992ARA&A..30..575C} are not entirely surprising given the different assumptions of each calibration.  

\citet{2005ApJ...631L..13D} report radio stacking (weighted average stacking + individual detections) of their $K_{Vega} < 20$ sample to obtain a luminosity of $3.6 \times 10^{23}$ W Hz$^{-1}$, corresponding to SFR $\sim$ 210 M$_\odot$ yr$^{-1}$, using the radio calibration of \citet{2001ApJ...554..803Y}, which is similar to our calibration of \citet{2003ApJ...586..794B}.   Our estimates from Table \ref{TABLE:RADIOSFR} are consistent with these results.
 
In stacking a K$_{AB} \leq$ 23 sample of {\it BzK}s, \citet{2009MNRAS.394....3D} reported  a median s{\it BzK} luminosity of 1.28 $\times$10$^{23}$ W Hz$^{-1}$ corresponding to SFR=154$\pm$7 M$_\odot$ yr$^{-1}$, which is similar to our 1.2$<z\leq$1.5 bin result of 138$\pm$11 M$_\odot$ yr$^{-1}$, although a formal comparison is not possible because of the 1.2 magnitude shallower depth of this present sample.  Likewise, the values presented here are similar to results from the COSMOS survey ($K_s<23$ selected sample; SFR in the range 30-100 M$_\odot$yr$^{-1}$ \citealt{2009ApJ...698L.116P}), where the radio-SFR calibration of \citet{2001ApJ...554..803Y} is used.
\subsection{X-Ray SFR Estimates}
\label{XraySFRSection}

Our {\it {\it Chandra}} soft band stacked X-ray fluxes are in the range 3.2-9.5 $\times 10^{-18}$ erg s$^{-1}$ cm$^{-2}$, see Table \ref{TABLE:XRAYSFR}.  Using observed frame, soft band fluxes and our assumed $\Gamma = 1.2$ spectrum to convert flux to luminosity leads to rest frame, 2-10 keV luminosities in the range $7-50 \times 10^{40}$ erg s$^{-1}$, indicated in Table \ref{TABLE:XRAYSFR}.  

In comparison with other reported values of galaxies detected to the same $K$ band depth as presented here,  \citet{2004ApJ...617..746D} find rest frame, 2-10 keV luminosity of $8.6\times10^{41}$ erg s$^{-1}$ (they use $\Gamma$ = 2.1 in their flux to luminosity conversion) in stacking 23 s{\it BzK}s in the K20 Survey \citep{2002A&A...381L..68C} that includes part of the CDF-S.  \citet{2005ApJ...631L..13D} find a rest frame, 2-10 keV luminosity of $3.4 \times 10^{41}$ erg s$^{-1}$ in stacking X-ray undetected s{\it BzK}s in 2 Ms {\it {\it Chandra}} data in GOODS-N (they use $\Gamma$ = 2.0 in their flux to luminosity conversion).   Thus we conclude that our stacked X-ray luminosities are consistent with other $z\sim2$ star-forming galaxies reported in the literature.

Our SFRs estimated from the rest frame 2-10 keV luminosities, and the calibration of \citet{2003A&A...399...39R} are in the range 15-100 M$_\odot$ yr$^{-1}$, a factor of $\approx 5$ lower than the corresponding calibrations of \citet{2010ApJ...724..559L} and \citet{2004A&A...419..849P}.  As discussed in Section \ref{X-RayLuminositySection}, SFR$_{2-10 keV}^{Persic}$ may be considered to provide upper limits.  In the SFR calibration of \citet{2010ApJ...724..559L}, the stellar mass term contributes $<$23\%, 7\%, 3\% and 1.4\% to the SFRs in each redshift range of Table \ref{Table:SFR}, respectively.  The trend of decreasing contribution from LMXBs to the total X-ray luminosity as redshift increases is broadly consistent with models of the LMXB population and star formation history that predict the LMXB population to decline at $z>1$ \citep{2001ApJ...559L..97G}.

\section{DISCUSSION}
\label{DiscussionSection}

\subsection{Comparison of SFR Estimates}

SFR estimates from X-ray through radio calibrations are compared in Table \ref{Table:SFR}.  The SFR$_{IR}$ values in Table \ref{Table:SFR} are obtained from L$_{IR}$ estimates from fits of MIR-radio ($\lambda \geq 24 \mu$m) photometry to CE01 templates.  To gauge their consistency, we plot the ratio of SFR computed from each calibration to SFR$_{IR+UV}$ in Figure \ref{FIG:SFRato}.  In this figure, SFR$_{IR}$/SFR$_{IR+UV}$ is most nearly equal to one, reflecting that IR luminosity accounts for $>$90\% of the total SFR in these galaxies.  

Comparing  SFR$_{24 \mu m}$ to SFR$_{IR+UV}$ in Table \ref{Table:SFR} and Figure \ref{FIG:SFRato} illustrates the overestimate of SFR at high redshift from this single waveband estimate.  The resulting poor fit to the data in the highest redshift bin, $2.0<z\leq3.2$, illustrated in Figure \ref{FIG:CE01TemplateFits}, overestimates L$_{IR}$ by approximately a factor of 6.  Although 24 $\mu$m estimates of L$_{IR}$ can be robust at $z<1.5$ (e.g. \citealt{2010A&A...518L..29E}), overestimation of L$_{IR}$, particularly at higher redshift has been previously reported (e.g. \citealt{2007ApJ...668...45P,2009ApJ...698.1380M,2010ApJ...725..742M,2010A&A...518L..24N,2011A&A...533A.119E}).  Using {\it Spitzer} IRS spectroscopy, \citet{2009ApJ...698.1380M} concluded that this luminosity over-estimate arises due to unusually large polycyclic aromatic hydrocarbon features in these galaxies, and to a lesser extent, AGN contamination.  

Dust-corrected UV SFRs, shown in Table \ref{Table:UVSFR}, from individual fits to s{\it BzK} photometry agree with SFR$_{IR+UV}$ in the middle redshift bin, although the highest redshift bin exhibits larger SFR$_{UV}^{Corr}$ and also larger error than lower redshift bins.  This large SFR value is strongly affected by a small number of outliers with very high computed SFRs, which may suggest a modified extinction law for at least some galaxies above $z \sim 2$.  Also, we find SFR$_{UV}^{Corr}$ to underestimate SFR$_{IR+UV}$ in the lowest redshift bin by a factor of five.  SFR$_{UV}^{Corr}$ values were computed using fits to the entire available broadband and medium band photometry (typically 12-14 sampled points in each fit); however, fits that were computed based upon broadband photometry only (typically only 3 measured points in each fit) produced systematically higher SFRs.  The medium band photometry clearly better samples the observed SED, and these results suggest a discrepancy between SFR$_{UV}^{Corr}$ and SFR$_{IR+UV}$ for {\it BzK}s in this redshift range.

Agreement between dust-corrected UV SFR and SFR$_{IR+UV}$ in {\it BzK}s to within a factor of $\sim$~2 has been reported previously (e.g. \citealt{2006ApJ...644..792R,2007ApJ...670..156D,2010A&A...518L..24N}).  \citet{2010ApJ...712.1070R} find SFR$_{UV}^{Corr}$ to agree with SFR determined from H$\alpha$ spectroscopy for LBGs at $z\sim 2$.  In particular, \citet{2006ApJ...644..792R} find LBGs and {\it BzK}s with ages $>$ 100 Myr to follow the \citet{1999ApJ...521...64M} relation while LBGs and {\it BzK}s with ages $<$ 100 Myr have rest UV colors that are redder than expected for a given L$_{IR+UV}$.  

We have tried two methods of estimating radio fluxes: weighted average stacking of non-detections averaged with individual detections and median stacking of all sources.  We find results of median stacking to be more consistent with SFR$_{IR+UV}$, despite the fact that SFR$_{IR+UV}$ estimates were obtained with the method of weighted average stacking.  Radio outliers in our sample may be due to residual AGN contamination.  

Among radio based SFR estimates, Figure \ref{FIG:SFRato} illustrates agreement between SFR$_{1.4 GHz}^{Bell}$ and SFR$_{IR+UV}$ to within a factor of two over our redshift range; this agreement is a consequence of the IR-Radio correlation for s{\it BzK}s, which is assumed in the calibration of SFR$_{1.4 GHz}^{Bell}$.  The model of \citet{1992ARA&A..30..575C} does estimate star formation from radio luminosity independent of the IR-Radio correlation; SFR$_{1.4 GHz}^{Condon}$ exceeds SFR$_{IR+UV}$ by a factor of two over the observed redshift range, but the ratio of these two calibrations appears relatively insensitive to redshift.  The discrepancies between the radio SFR calibrations of \citet{2003ApJ...586..794B} and \citet{1992ARA&A..30..575C} are not entirely surprising given the different assumptions of each calibration. 


X-ray SFR indicators show wide variation in their ratios to SFR$_{IR+UV}$.  As discussed in Section \ref{XraySFRSection}, SFR$_{2-10 keV}^{Persic}$ may be interpreted as an upper limit.   SFR$_{2-10 keV}^{Lehmer}$ yields  estimates that are similar to SFR$_{2-10 keV}^{Persic}$ because of the subdominant contribution of the stellar mass term to the X-ray luminosity in these galaxies.  Meanwhile SFR$_{2-10 keV}^{Ranalli}$ actually agrees with SFR$_{IR+UV}$ to within a factor of two for $z>1.5$.  SFR$_{2-10 keV}^{Ranalli}$ has been applied to {\it BzK}s (e.g. \citealt{2007ApJ...670..173D,2005ApJ...633..748R}), BX/BM galaxies (e.g. \citealt{2004ApJ...600..695R}), and LBGs at $z\sim 2$ (e.g. \citealt{2010ApJ...712.1070R}).  In these studies, X-ray SFRs often agree with other waveband estimators, typically to within the same factors as reported here.  

However, interpretation of these X-ray SFRs depends upon an uncertain contamination fraction from obscured AGN.  We speculate that AGN contamination may be present in our sample, particularly in the highest redshift bin, in which we find L$_{2-10 keV} \sim 10^{42}$ erg s$^{-1}$.  AGN contamination even at the level of 10\% can require downward adjustment to X-ray SFRs by a factor of 2-5 \citep{2008ApJ...681.1163L}.  Consequently, X-ray SFRs would be over-estimated.  \citet{2008ApJ...681.1163L} compute a luminosity dependent AGN fraction in order  to correct X-ray stacking results in $z\sim 3$ galaxies; they find that $\approx 50-70\%$ of the stacked 0.5-2 keV counts may arise from obscured AGN.  If a similar AGN fraction exists in our sample, then the X-ray SFRs would need to be adjusted downwards by a factor of $\sim$2.5, bringing SFR$_{2-10 keV}^{Persic}$ and SFR$_{2-10 keV}^{Lehmer}$ into better agreement with other waveband indicators, and taking SFR$_{2-10 keV}^{Ranalli}$ out of agreement with other waveband indicators.  

\subsection{Sources of Uncertainty}

In computing uncertainties in these SFR estimates,  we have considered the effects of errors in photometry, spectral shape and redshift on luminosity estimates.  These errors are reported in Table \ref{Table:SFR} and indicate a wide range in precision of the various waveband indicators.  IR estimates are the most precise; X-ray SFR calibrations are the least precise because they rely upon empirical correlations with IR luminosity and thus introduce additional scatter into the SFR estimate.  Each of these SFR calibrations assume continuous star formation of at least 10$^8$ Myr and solar metallicity; therefore different assumptions about timescales or chemical evolution cannot account for systematic differences between the calibrations.

We have excluded the uncertainty due to luminosity$-$SFR calibration.  We are not able to assess the absolute uncertainties in these SFR calibrations because in many cases they contain implicit dependencies on $L_{IR}$--SFR calibration.   SFR$_{IR}$, SFR$_{24 \mu m}$, SFR$^{Bell}_{1.4 GHz}$, SFR$^{Ranalli}_{2-10 keV}$ and SFR$^{Persic}_{2-10 keV}$ depend implicitly upon the $L_{IR}$--SFR calibration of \citet{1998ApJ...498..541K}, which has a reported systematic uncertainty of about a factor of 2-3.  Similarly, SFR$^{Lehmer}_{2-10 keV}$ depends upon the $L_{IR}$--SFR calibration of \citet{2005ApJ...625...23B}, which also has a reported systematic uncertainty of a factor of 2.  Similarly, SFR$^{Corr}_{UV}$ depends upon a model dependent UV luminosity -- SFR calibration, for which various published values may differ by a factor of 2 \citep{1998ARA&A..36..189K}. In the comparisons discussed here, we assume that the various SFR calibrations are consistent with each other and do not evolve with redshift.  We compare them against each other, and disagreement can provide evidence of systematic offsets in a given calibration.   

For generalizing the results of this sample of {\it BzK}s to the general population of {\it BzK}s, errors may be estimated from bootstrap resampling, which would incorporate sample variance, and undoubtedly increase the size of random errors.  We do not consider this sample variance here, because our primary aim is to compare the SFR calibrations to each other, to assess their consistency, rather than comparing SFRs of {\it BzK}s as a population to other populations of star-forming galaxies.

We consider the differing sensitivities to star formation among the various wavebands.  We estimate the lowest average SFR detectable in each waveband and redshift bin from the reported image depths by computing the average flux that would yield a 3$\sigma$ stacked detection given our sample sizes.  Converting this flux to a minimum detectable average luminosity at each bin median redshift yields SFR values $<3, \sim3$, and $\sim15$ M$_\odot$yr$^{-1}$ for SFR$^{Uncorr}_{UV}$, SFR$_{24 \mu m}$, and SFR$_{Radio}$ respectively, at all sampled redshifts.  SFR$^{Corr}_{UV}$ and SFR$_{IR}$ are even more sensitive than SFR$^{Uncorr}_{UV}$ and SFR$_{24 \mu m}$; quantitative estimate would require detailed modeling of the SED fitting procedure.  The X-ray calibrations are less sensitive, with minimum detectable SFR in the range 30-66 and 150-330 M$_\odot$yr$^{-1}$ (increasing with redshift) for SFR$^{Ranalli}_{2-10 keV}$ and SFR$^{Persic}_{2-10 keV}$ respectively.  X-ray SFR estimators have low sensitivity and are subject to bias due to residual AGN contamination and it is not clear which of these effects explains the overestimation of SFR.

\section{CONCLUSION}
\label{ConclusionSection}
The main results of this paper are summarized in the comparison of SFR indicators given by Table \ref{Table:SFR} and Figure \ref{FIG:SFRato}.  We consider SFR determined from panchromatic estimation of L$_{IR}$ the most comprehensive approach, where the available data exist, notwithstanding the challenges of accurate fitting of spectral templates.  With this method, average SFR of redshift binned galaxies can be determined with $\sim$5\%$-$10\% random uncertainty (though considerably larger for individual sources); however, the total error is dominated by a factor of $\sim$2 systematic uncertainty.

This systematic uncertainty, not included in the errors indicated here, of a factor of $\sim 2$ is present in the $L_{IR}$$-$SFR and $L_{UV}$$-$SFR calibrations that underlie each method of SFR estimation.  Each of the calibrations discussed in this paper either implicitly or explicitly assumes a continuous star formation history and solar metallicity; consequently differing assumptions about these parameters should not account for systematic differences.

We find that dust-corrected UV SFR, using the method of \citet{1999ApJ...521...64M}, agrees with SFR$_{IR+UV}$ for galaxies in the range $1.2 < z \leq 1.5$, but over-estimates SFR$_{IR+UV}$ in the range $2.0<z\leq3.2$ due to a small population of outliers.  Radio SFRs estimated from the calibration of \citet{2003ApJ...586..794B} are in agreement with the total SFRs to within errors.  The SFR calibration of \citet{1992ARA&A..30..575C} overestimates the total SFR by a factor of two for {\it BzK} galaxies over this redshift range.

Perhaps surprisingly, the calibration of \citet{2003A&A...399...39R} yields SFR estimates that agree with SFR$_{IR+UV}$ and other indicators in the range $1.5 < z \leq 3.2$, while the better suited X-ray calibrations of \citet{2004A&A...419..849P} and \citet{2010ApJ...724..559L} over-estimate SFR in the range $1.5<z<3.2$.  One possible explanation is mild AGN contamination in the upper two bins, which would bring SFR$_{2-10 keV}^{Persic}$ and SFR$_{2-10 keV}^{Lehmer}$ into rough agreement with SFR$_{IR+UV}$ and cause SFR$_{2-10 keV}^{Ranalli}$ to underestimate SFR.  X-ray SFR estimates are also notably less sensitive than other waveband indicators.

There is an evident variation in the accuracy and precision of SFR calibrations, and each method has its limitations and caveats. Radio and X-ray SFR calibrations rely upon empirical correlations of flux in their wavebands to $L_{IR}$ that introduce inevitable scatter, particularly in the X-ray calibrations.  24 $\mu$m only and UV SFRs work for most galaxies but have important exceptions and cannot be applied universally.  

Our analysis of $K_{AB} < 21.8$ s{\it BzK}s in the redshift range $1.2<z\leq3.2$ confirms that they are IR luminous, star-forming galaxies for which approximately 90\% of the total star formation is obscured by dust.  By fitting to CE01 templates, we find average IR luminosities for redshift binned s{\it BzK} galaxies at median redshifts 1.4, 1.8, 2.2 to be  $3.0\pm0.3 \times 10^{11}$ L$_\odot$, $4.0\pm 0.4 \times 10^{11}$ L$_\odot$, $8.3\pm0.7 \times 10^{11}$ L$_\odot$.  We find SFR$_{IR+UV}$ at these redshifts to be 55$\pm$6, 74$\pm$8, 154$\pm$17 M$_\odot$ yr$^{-1}$ respectively.

\acknowledgments

We acknowledge valuable conversations with and comments by Viviana Acquaviva, Eric Bell, Nicholas Bond, Lucia Guaita, Bret Lehmer, Felipe Menanteau, and Axel Weiss and we thank LESS for providing the 870 $\mu$m data, obtained from LABOCA APEX RUN IDs: 078.F-9028(A), 079.F-9500(A), 080.A-3023(A), and 081.F-9500(A).  Support for this work was provided by the National Science Foundation under grants AST-0807570 and AST-1055919 and by NASA through an award issued by JPL/Caltech.  EG thanks U.C. Davis for hospitality during the preparation of this manuscript.  Support for the work of ET was provided by the National Aeronautics and Space Administration through Chandra Postdoctoral Fellowship Award Number PF8-90055 issued by the {\it Chandra} X-ray Observatory Center, which is operated by the Smithsonian Astrophysical Observatory on behalf of the National Aeronautics and Space Administration under contract NAS8-03060.  
\section{\bf APPENDIX:  ERROR ESTIMATION IN STACKING} 
\label{Appendix:AggregateFluxEstimation}

\subsection{Ordinary Average}

Section \ref{MethodSection} presents the method used in this paper for estimating the aggregate flux, $\mu$,  from $N$ prior positions that include a combination of $I$ individual detections, x$_i$, and a stacking detection, x$_S$, from $S$ undetected sources, where $N=I+S$, given by
\begin{equation}
\mu = \frac{1}{N} \left( \sum_{i=1}^I{x_i} + S x_s \right)
\label{EQ:UnWeightedAverage}
\end{equation}
The variance  $\sigma_\mu^2$ of this estimate is related to the individual errors $\sigma_i$ and $\sigma_s$ as well as their covariances, $\sigma_{i,s}$ from standard error analysis
\begin{equation}
\sigma_\mu^2 = \sum_{i=1}^I{\left( \frac{\partial \mu}{\partial x_i} \right)^2\sigma_{x_i}^2}+  \left( \frac{\partial \mu}{\partial x_s} \right)^2\sigma_{x_s}^2+ \sum_{i=1}^{I}{2 \sigma_{i,s}^2}  \frac{\partial \mu}{\partial x_i}\frac {\partial \mu}{\partial x_s} + ...
\end{equation}
We assume that the covariances between individual detections and the stacking detection, $\sigma_{i,s}$ and the covariances between separate individual detections, $\sigma_{i,j}$, are zero. 
\begin{equation}
\sigma_\mu^2 = \sum_{i=1}^I{\left( \frac{\partial \mu}{\partial x_i} \right)^2\sigma_{x_i}^2}+  \left( \frac{\partial \mu}{\partial x_s} \right)^2\sigma_{x_s}^2
\end{equation}
\begin{equation}
\sigma_\mu^2 = \frac{1}{N^2} \sum_{i=1}^I \sigma_{x_i}^2 + \left( \frac{S}{N} \right)^2 \sigma_{x_s}^2
\end{equation}
\begin{equation}
\sigma_\mu = \frac{1}{N} \sqrt{\sum_{i=1}^I \sigma_{x_i}^2 + S^2 \sigma_{x_s}^2}
\label{EQ:UnweightedAverageError}
\end{equation}
Equations \ref{EQ:UnWeightedAverage} and \ref{EQ:UnweightedAverageError} are used to compute the average and error, respectively, of individual and stacking detections reported in this paper.

In the case of all measurement errors being equal, which is a good approximation in the case of LESS data, 
then $\sigma_i \equiv \sigma$ and $\sigma_s = \sigma / \sqrt{N}$.  Using Equation \ref{EQ:UnweightedAverageError}, $\sigma_\mu$ is computed as
\begin{equation}
\sigma_\mu = \frac{1}{N} \sqrt{\sum_{i=1}^I \sigma^2 + S^2 \frac{\sigma^2}{S}}
\end{equation}
\begin{equation}
\sigma_\mu = \frac{1}{I+S} \sqrt{I \sigma^2 + S\sigma^2}
\end{equation}
\begin{equation}
\sigma_\mu = \frac{1}{\sqrt{I+S}} \sigma
\end{equation}
\subsection{Weighted average:}  An alternative approach to combining individual and stacking detections into a single aggregate flux estimate is to use a weighted average of individual and stacking detections.  Assume there are $N=I+1$ flux measurements consisting of $I$ individual detections and a single stacking detection.  Each individual flux measurement and the stacked flux measurement is considered as an independent flux measurement for the purpose of computing an average, and these measurements are combined as an inverse variance weighted average.
\begin{equation}
\mu^\prime= \frac{\sum_{i=1}^N{\frac{x_i}{\sigma_i^2}}}{\sum_{i=1}^N{\frac{1}{\sigma_i^2}}}
\label{EQ:WeightedAverage}
\end{equation}
\begin{equation}
\sigma_{\mu^\prime}^2 = \frac{1}{\sum{\frac{1}{\sigma_i^2}}}
\label{EQ:WeightedAverageError}
\end{equation}
Equations \ref{EQ:WeightedAverage} and \ref{EQ:WeightedAverageError} are also used to estimate the aggregate flux and error for representative data reported in this paper.  In particular, the errors computed according to this method are numerically equal to the errors computed from the ordinary average, Equation \ref{EQ:UnweightedAverageError}, to better than three significant digits.   

For clarity, Equation \ref{EQ:WeightedAverageError} can be written to include stacking and individual detections separately
\begin{equation}
\sigma_{\mu^\prime}^2 = \frac{1}{\sum_{i=1}^{I}{\frac{1}{\sigma_i^2} + \frac{1}{\sigma_s^2}}}
\label{EQ:WeightedAverageErrorSeparate}
\end{equation}
In the case of all measurement errors being equal, Equation \ref{EQ:WeightedAverageErrorSeparate} can be simplified using $\sigma_i \equiv \sigma$ and $\sigma_s = \sigma / \sqrt{N}$, just as in the ordinary average error computation:
\begin{equation}
\sigma_{\mu^\prime}^2 = \frac{1}{\sum_{i=1}^{I}{\frac{1}{\sigma^2} + \frac{S}{\sigma^2}}}
\end{equation}
\begin{equation}
\sigma_{\mu^\prime}^2 = \frac{\sigma^2}{I+S}
\end{equation}
\begin{equation}
\sigma_{\mu^\prime} = \frac{\sigma}{\sqrt{I+S}}
\end{equation}
Thus for the case of all measurement errors being equal, the error of the weighted average is identical to the error of the ordinary average. 

\subsection{Comparison of ordinary and weighted average:}

In the case of identical errors for the individual measurements, $\sigma_i \equiv \sigma$, then weighted and ordinary averages give the same result for the aggregate flux.   In the case of measurement errors being unequal, then the weighted average will in principle have the smaller error; however, the difference will be small, and if the errors are not independent, e.g. if brighter sources have larger errors, then the weighted average introduces a bias to the flux estimate $\mu^\prime$.  For instance, if dim sources are always measured with better precision than bright sources, then a weighted average of the population of all sources will always be biased toward dim sources.  This circumstance could arise if flux measurement errors are dominated by Poisson counting statistics.  However, if the flux measurement errors are uncorrelated with the flux, then there will be no problem with the weighted average.  

\bibliographystyle{apj}                       

\bibliography{pkurczynski_SFRCompare}

\clearpage

%
\begin{figure}[h]
\begin{center}

\includegraphics[scale=0.5]{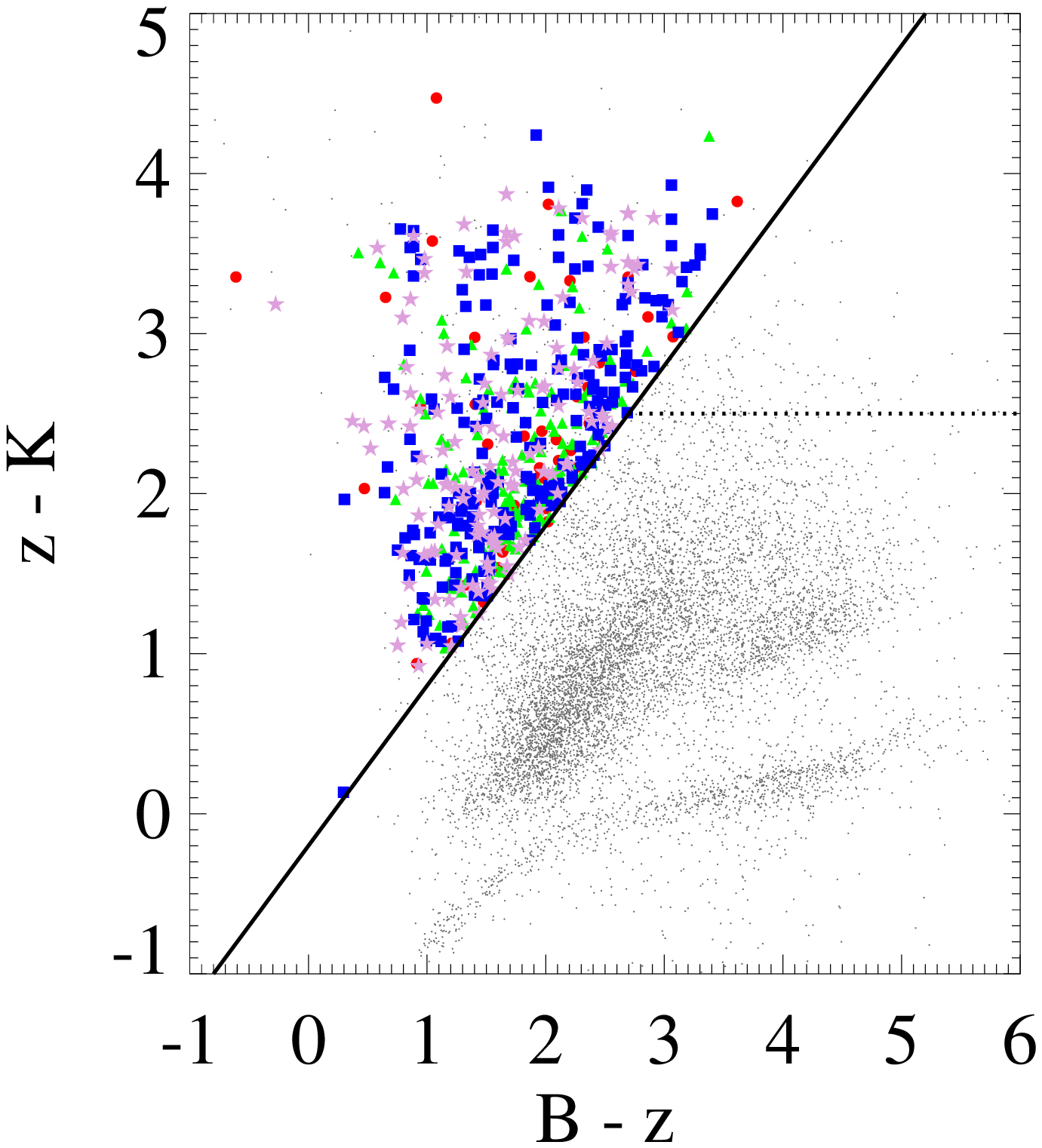}

\vspace{-0.5cm}
\end{center}
\caption{
{\it BzK} diagram illustrating corrected NIR-optical colors $(z-K)_{AB}$ vs. $(B-z)_{AB}$ for MUSYC $K<21.8$ selected sources (gray points), including redshift binned, star-forming galaxies (with X-ray detected AGN removed, see the text) in the range $0.9<z\leq1.2$ (red circles), $1.2<z\leq1.5$ (green triangles), $1.5<z\leq2.0$ (blue squares) and $2.0<z\leq3.2$ (purple stars).  The $sBzK$ region is located above the diagonal line, and is defined as $BzK\geq0.2$.  The $pBzK$ region is the wedge shaped area above the horizontal dotted line, $z-K > 2.5$. }

\label{FIG:BzKDiagram}
\end{figure}

%
%
\begin{figure}[h]
\begin{center}

\includegraphics[scale=0.5]{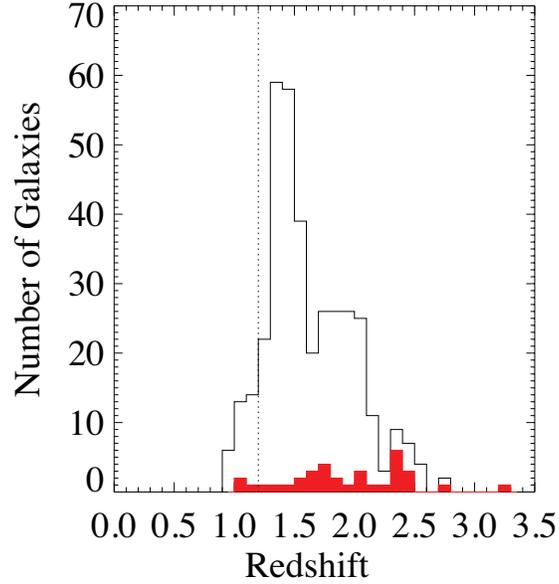}

\vspace{-0.5cm}
\end{center}
\caption{
Histograms of photometric and spectroscopic redshifts for s{\it BzK} galaxies in ECDF-S.  The upper histogram indicates photometric redshifts.  The lower filled (red), histogram indicates spectroscopic redshifts.  The vertical dotted line demarcates $z<1.2$ galaxies that are excluded from analysis.  
}
\label{FIG:ZHistograms}
\end{figure}

\clearpage

%
\begin{figure}[h]
\begin{center}

\includegraphics[scale=0.40]{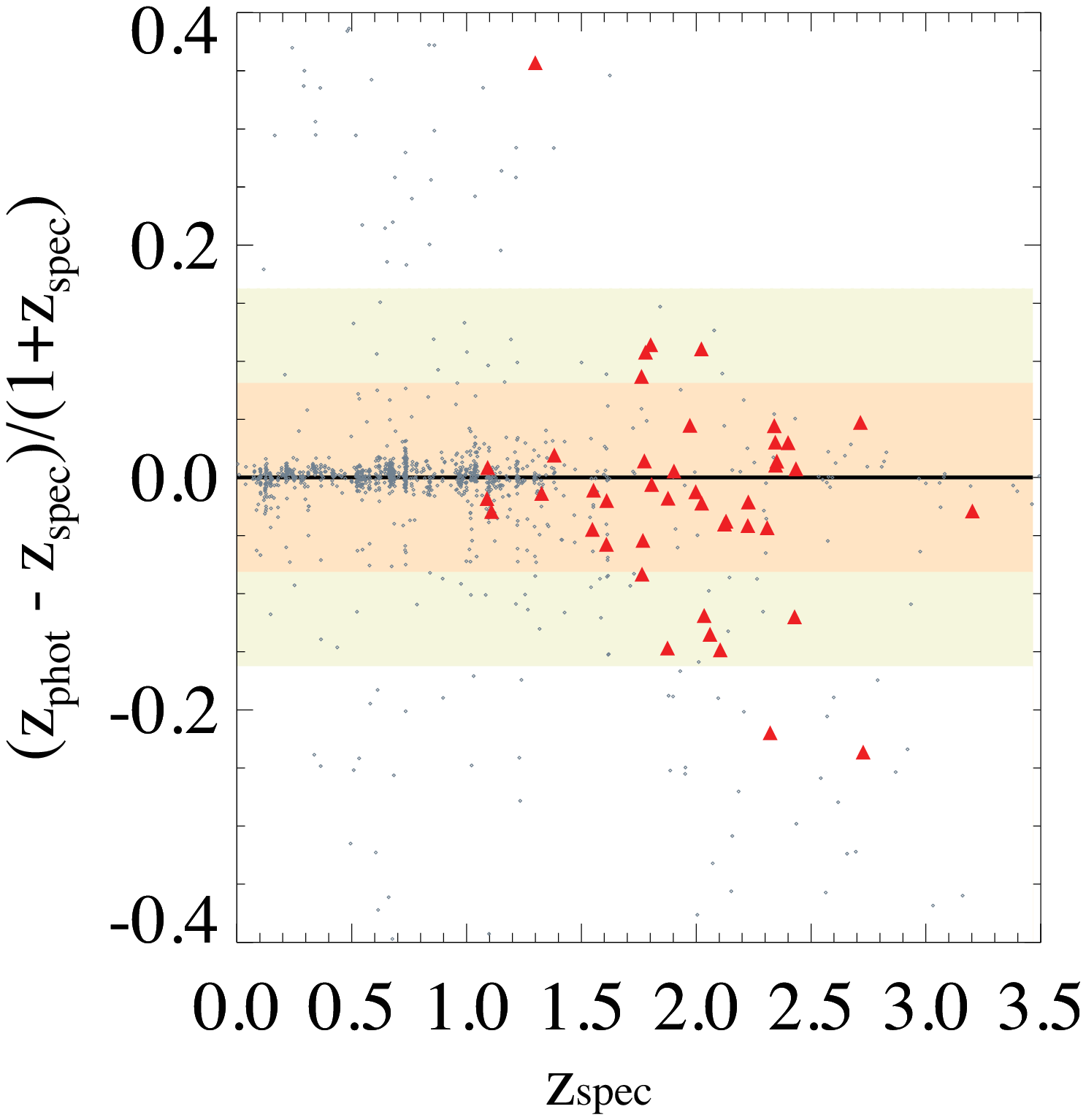}
\includegraphics[scale=0.40]{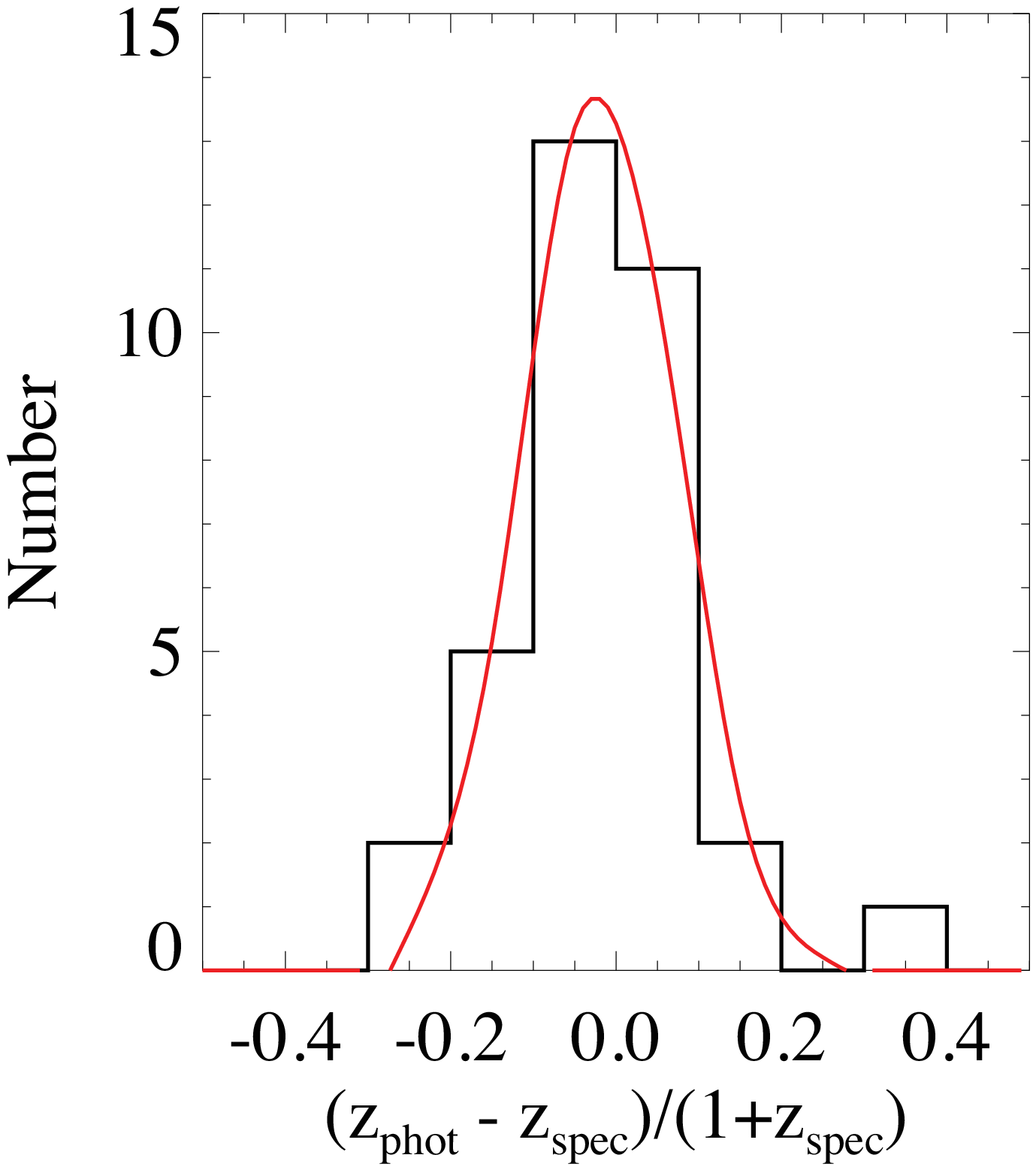}

\vspace{-0.5cm}
\end{center}
\caption{
Left panel illustrates photometric redshift error, defined as $(z_{phot} - z_{spec})/(1+z_{spec}$), vs. spectroscopic redshift for $K$ selected galaxies in ECDF-S.  Triangles (red) illustrate s{\it BzK} galaxies.  1$\sigma$ and 2$\sigma$ regions, determined from the fit to the histogram (right panel) are indicated by shading.  Points (gray) illustrate all 1285 $K$ selected galaxies for which spectroscopic and photometric redshifts are available.  The right panel illustrates histogram of photometric redshift errors for s{\it BzK} galaxies along with a Gaussian fit (mean = -0.02 and $\sigma$ = 0.09).
}
\label{FIG:PhotozVsSpecz}
\end{figure}

\clearpage

%
%
%
\begin{figure}[h]
\begin{center}
\includegraphics[scale=0.35]{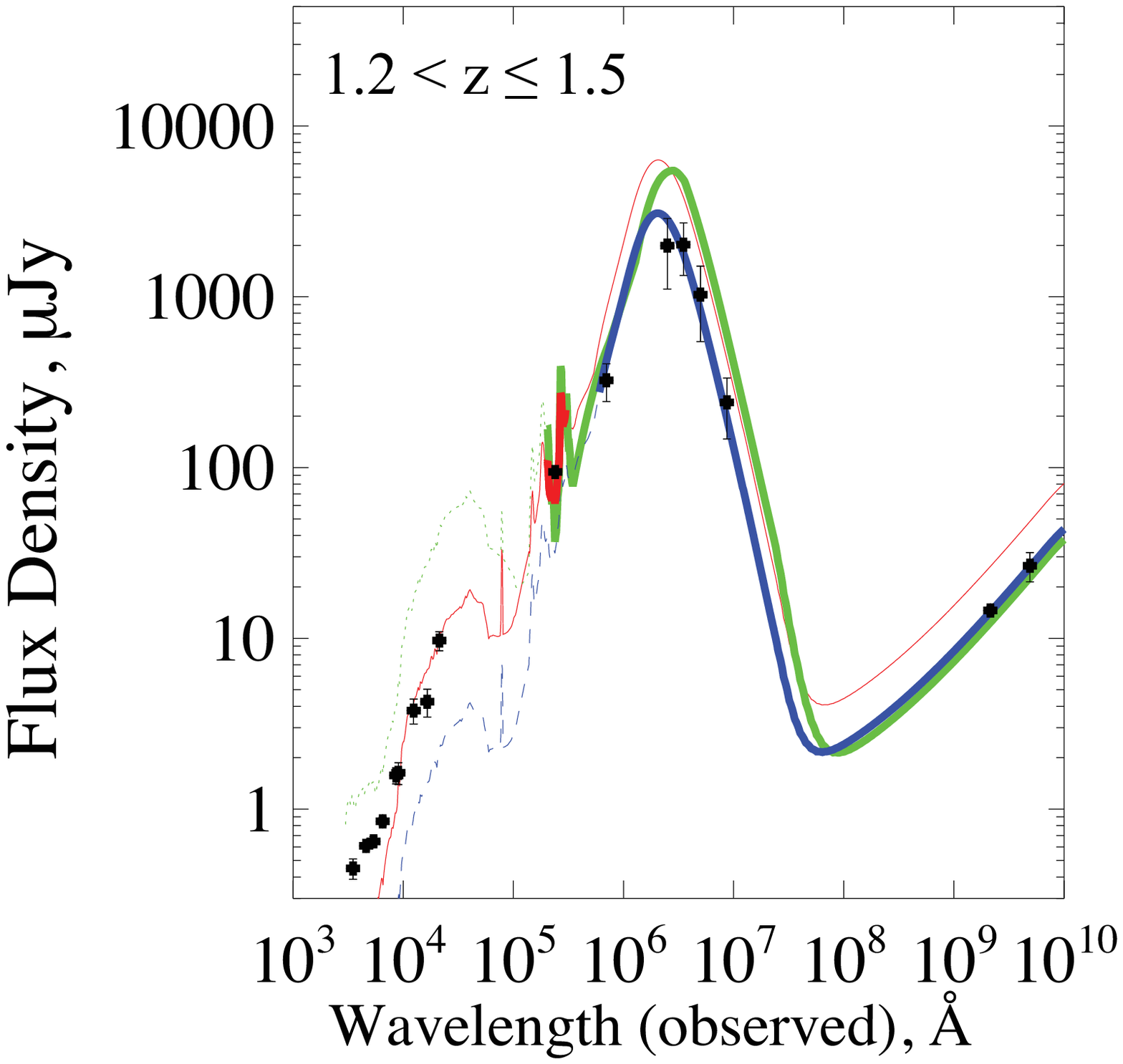} \\
\includegraphics[scale=0.35]{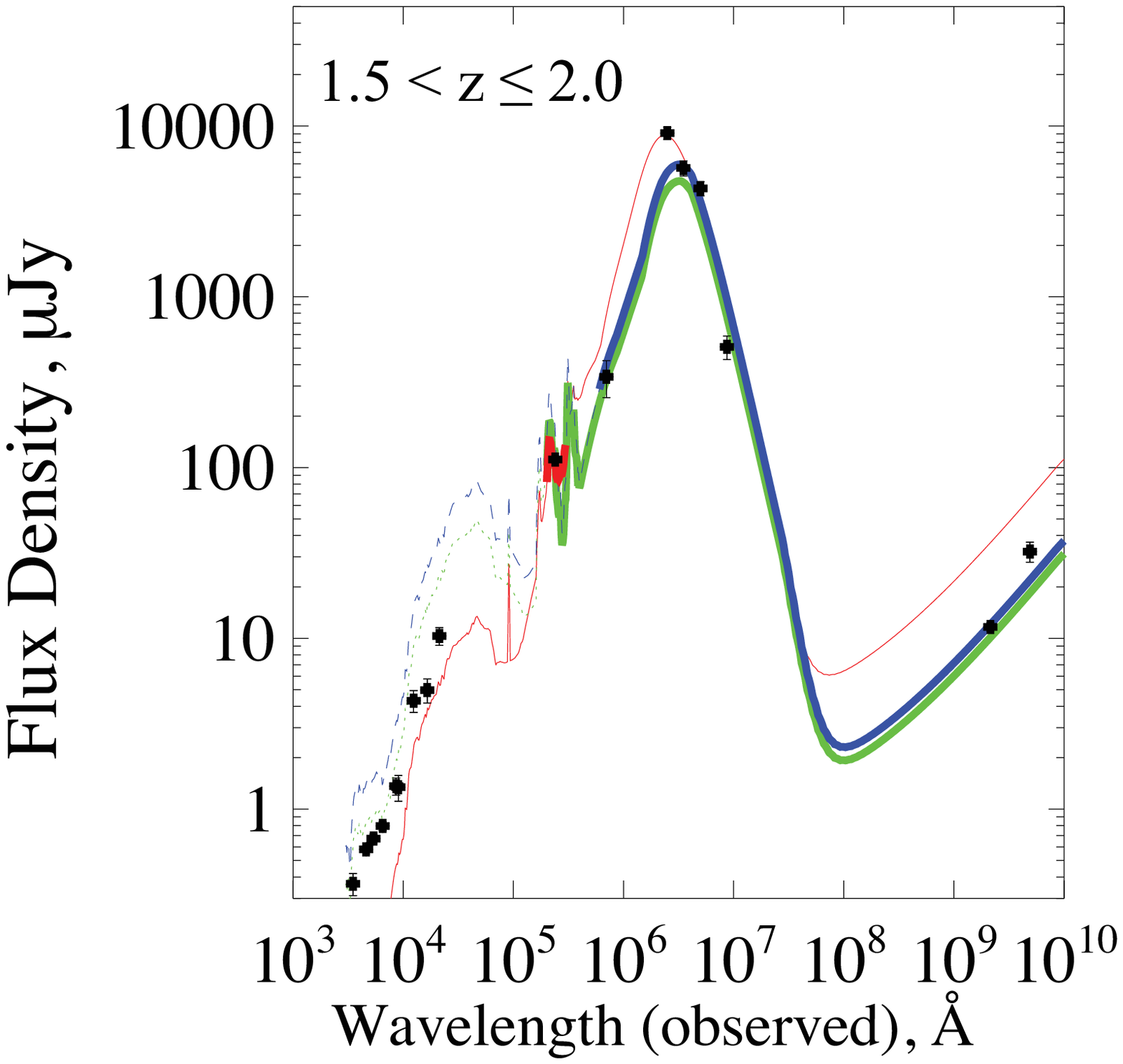} \\
\includegraphics[scale=0.35]{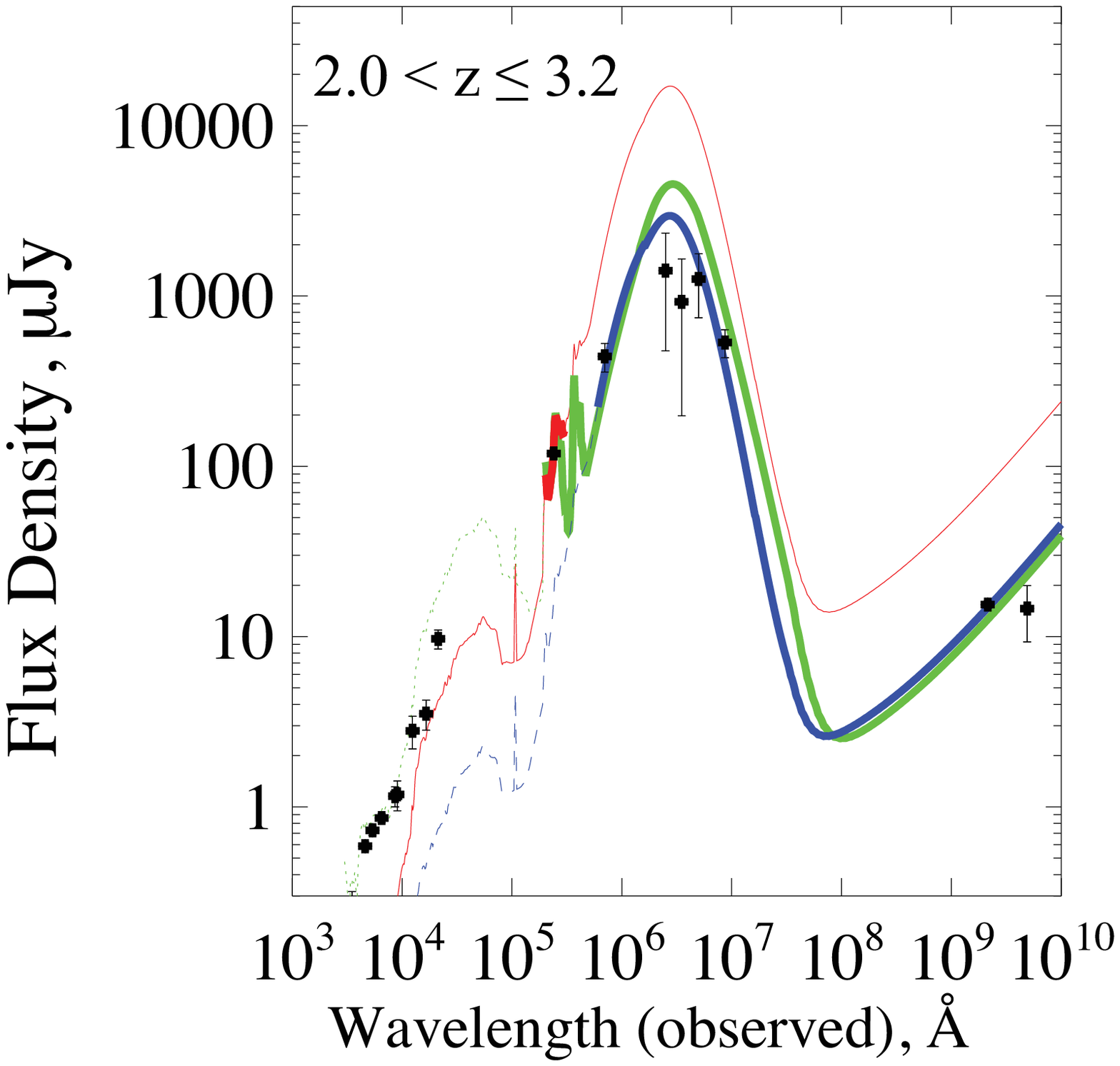} 
\vspace{-0.5cm}
\end{center}
\caption{Spectral energy distributions of redshift binned s{\it BzK} galaxies obtained from stacking analysis, and fits to various models. Spectral flux density, in units of $\mu$Jy, is plotted vs. observed frame wavelength, in units of Angstroms.  UV through radio flux measurements are indicated by points with error bars.  Curves indicate best fit models:  dotted (green) CE01 template fitted to the $\lambda \geq 24 \mu m$ spectrum, dashed (blue) CE01 template fitted to the $\lambda > 24 \mu m$ spectrum, and solid (red) CE01 template fitted to only the 24 $\mu$m data point.  For each model fit, the region of the spectrum used for the fit is indicated by the solid portion of the curve.  Top:   galaxies in the redshift range  $1.2<z\leq1.5$; middle $1.5<z\leq2.0$; bottom  $2.0<z\leq3.2$.  See also Table \ref{TABLE:IRFITSUMMARY} for fit details.}
\label{FIG:CE01TemplateFits}
\end{figure}

\clearpage

%
%
%
\begin{figure}[h]
\begin{center}

\includegraphics [scale=0.4]{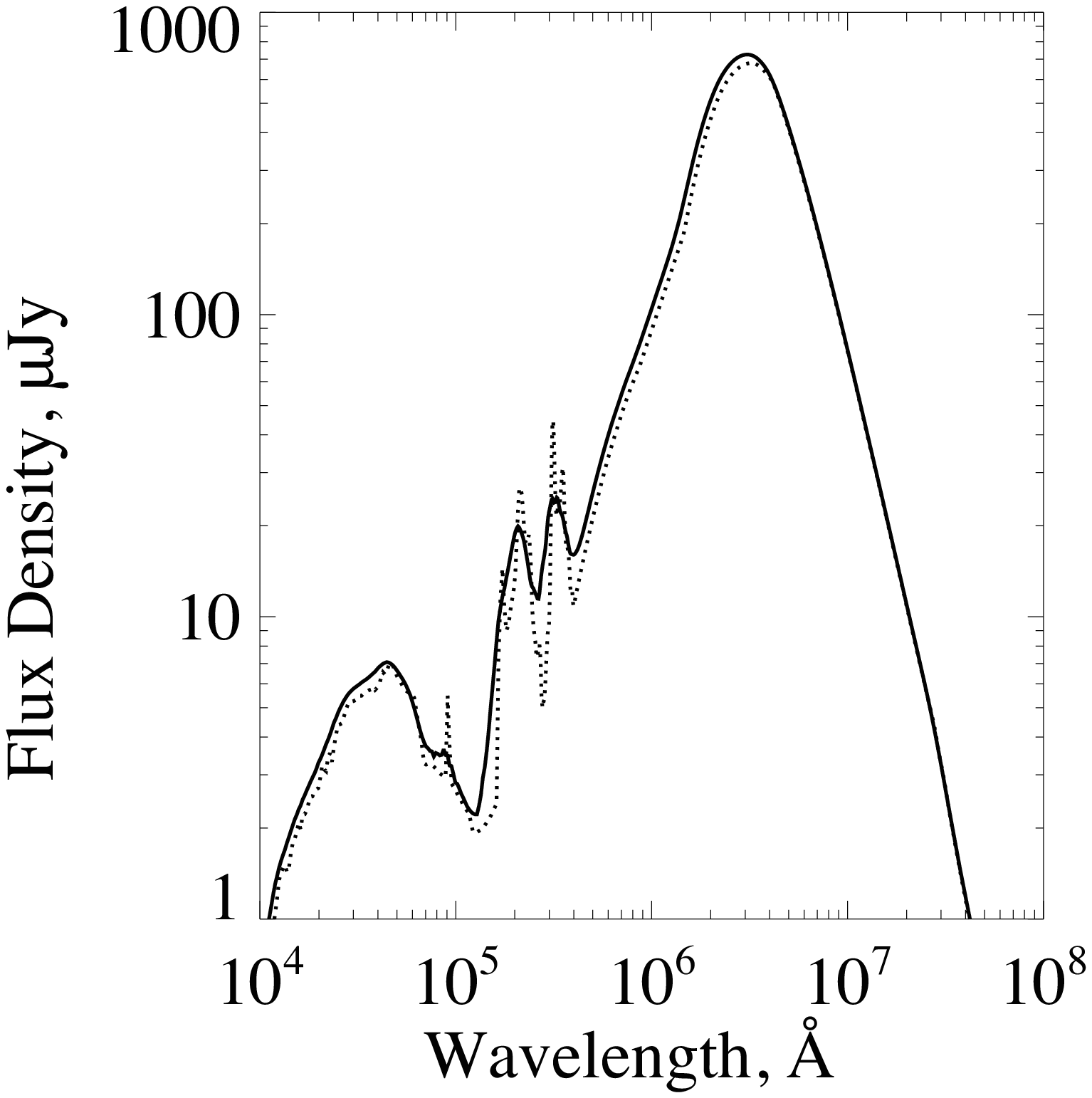}
\includegraphics [scale=0.4]{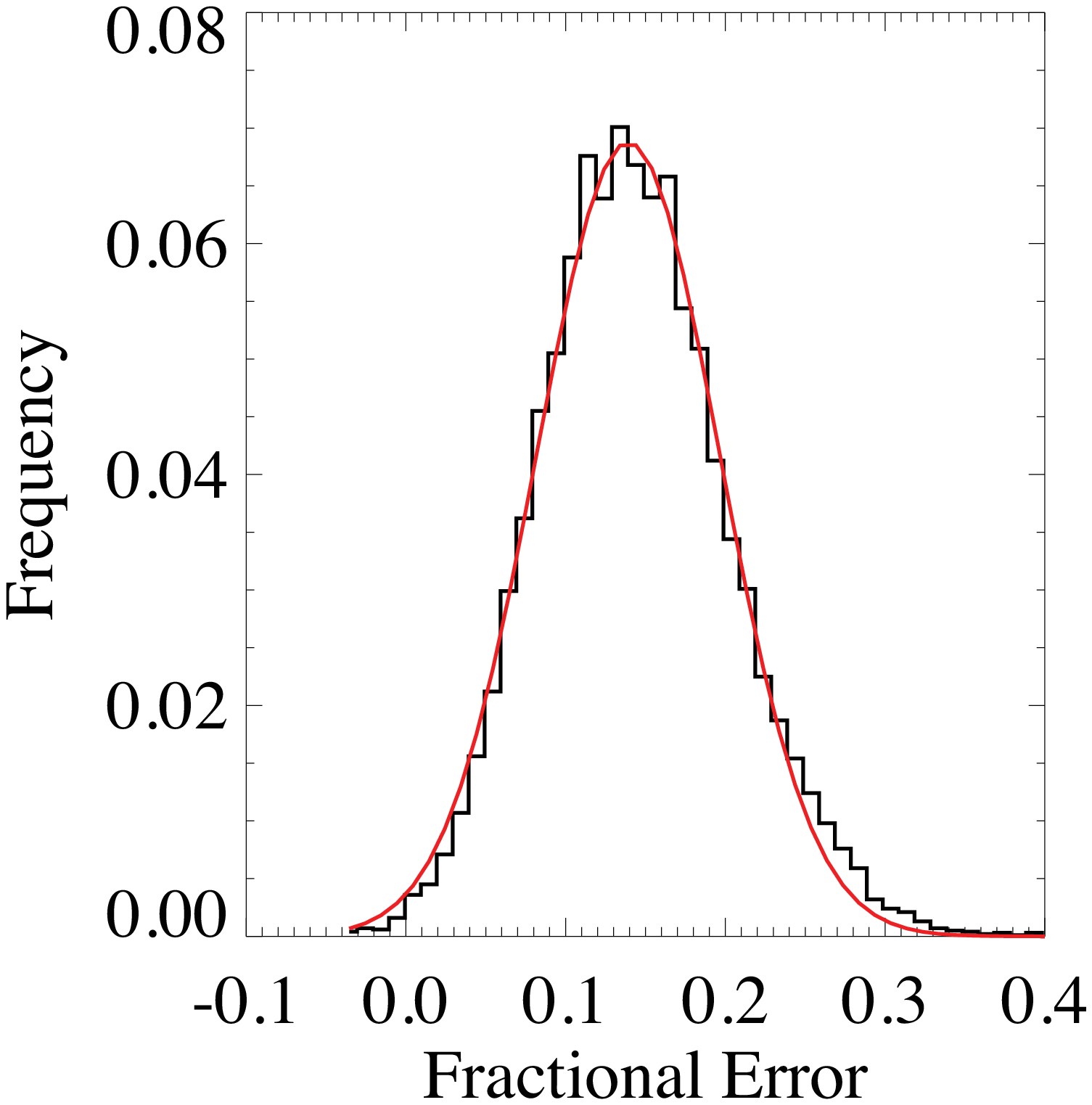}

\vspace{-0.5cm}
\end{center}
\caption{
Left panel illustrates simulated, observed frame spectra for s{\it BzK} galaxies in the range $1.5 < z \leq 2.0$.  The dotted curve indicates the best fit CE01 template placed at the median redshift, $z=1.756$.  The solid curve indicates the result of averaging 233 identical CE01 template sources, distributed in redshift according to the observed distribution of s{\it BzK}s, with photometric redshift errors added according to the distribution shown in Figure \ref{FIG:PhotozVsSpecz}.  The right panel illustrates the frequency distribution of L$_{IR}$ fractional errors due to photometric redshift errors and dispersion from individual galaxy SEDs, for s{\it BzK} galaxies in the range $1.5<z\leq 2.0$.  Solid curve (red) indicates a Gaussian fit with mean = 0.14 and $\sigma$=0.06.  Thus L$_{IR}$ for this redshift bin is overestimated by 14\%; photometric redshift and galaxy SED dispersion contribute a scatter of 6\%.
}

\label{FIG:ZBINSIMULATION}
\end{figure}


\clearpage

%
%
\begin{figure}[h]
\begin{center}
\includegraphics[scale=0.50]{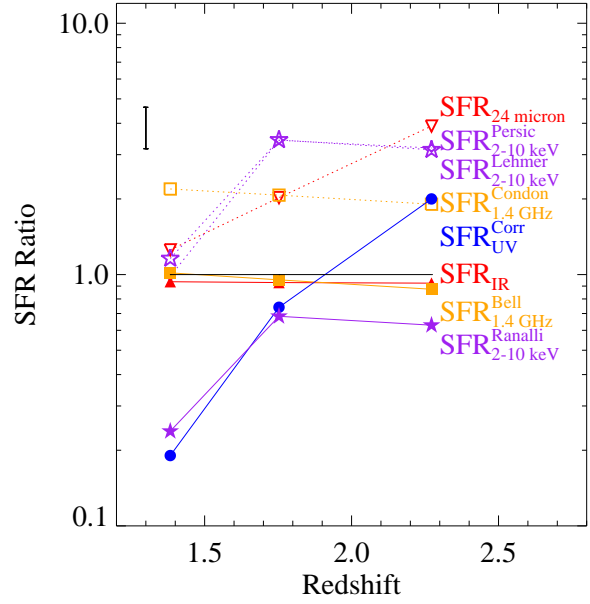}
\vspace{-0.5cm}
\end{center}
\caption{Ratio of SFR computed with various calibrations to the bolometric SFR estimate,  SFR$_{IR+UV}$.  Symbols are color coded by waveband:  X-ray (purple), UV (blue), IR (red), and radio (orange).  Lines connecting points are a guide to the eye.  A typical error bar for non-X-ray indicators is illustrated at left.  X-ray indicators have larger errors by a factor of $\sim$2.}
\label{FIG:SFRato}
\end{figure}


\clearpage

%
%
\begin{deluxetable}{cccccc}
\tablecolumns{6}
\tablecaption{Summary of Results of AGN Selection Criteria Applied to s{\it BzK} Sample.} 
\tablehead{
\colhead{Test} &
\colhead{Detected} &
\colhead{AGN} &
\colhead{X-ray} &
\colhead{q$_{IR}$} &
\colhead{IRAC} \\
\colhead{(1)}&
\colhead{(2)} & 
\colhead{(3)} & 
\colhead{(4)} & 
\colhead{(5)} & 
\colhead{(6)} 
}
\startdata
X-ray	&	110	&	107	&	{\bf 83}	&	7	&	18 	\\
q$_{IR}$   &	35	&	22	&	7	&	{\bf 15}	&	2  	\\
IRAC	&	649	&	25	&	18	&	2		&	{\bf 7}\\
\enddata
\tablecomments{Column (1) indicates the waveband test used for AGN discrimination.  Column (2) indicates the number of sources with data in each respective waveband.  Column (3) indicates the number of AGN confirmed by each waveband test irrespective of tests in other wavebands.  Entries in Columns (4-6) indicate the number of sources classified as AGN according to each waveband as follows:  Diagonal entries (in boldface) indicate sources that are uniquely identified as AGN in only one waveband.  Off diagonal entries indicate sources that are identified as AGN in at least the two wavebands indicated by the respective row and column headings.}
\label{TABLE:AGNCRITERIA}
\end{deluxetable}

\clearpage

%
%
%
%
\begin{deluxetable}{cclcccccc} 
\tablecolumns{9} 
\tablewidth{0pc} 
\tablecaption{Redshift Binned Sample of Non-AGN Star-Forming Galaxies (s{\it BzK}s) in the ECDF-S} 

\tablehead{
\colhead{} & \multicolumn{2}{c}{Total Sample}  & \colhead{ } & \multicolumn{5}{c}{Individual Detections} \\ 
\cline{2-3} \cline{5-9} \\ 
\colhead{Redshift} & \colhead{Median} & \colhead{Number} & \colhead{} & \colhead{MIPS} & \colhead{MIPS} & \colhead{LESS} & \colhead{VLA} & \colhead{{\it Chandra}} \\
\colhead{Range}  & \colhead{Redshift}   & \colhead{($z_{phot}, z_{spec}$)} & \colhead{} & \colhead{24 $\mu$m} & \colhead{70 $\mu$m} & \colhead{870 $\mu$m} & \colhead{1.4 GHz} & \colhead{0.5-2 keV} \\
\colhead{(1)} & \colhead{(2)}  & \colhead{(3)} & \colhead{} & \colhead{(4)} & \colhead{(5)}  & \colhead{(6)} & \colhead{(7)} & \colhead{(8)}  }
 
\startdata
1.2 $<z\leq$ 1.5 &  1.383 & 156 (152,4) 	&& 64 & 4 & 2 & 2 & 1  \\
1.5 $<z\leq$ 2.0 & 1.753  & 215 (202,13)  && 90 & 9 & 6 & 9 & 1 \\
2.0 $<z\leq$ 3.2 & 2.272  & 139 (122,17)  && 61 & 1 & 1 & 5 & 0 \\
\enddata

\tablecomments{Redshift binning scheme selected for this analysis.  Column (1) indicates the redshift range for each bin.  Column (2) indicates the median redshift of the sampled galaxies in each bin.  Column (3) shows the total number of galaxies in each bin, with numbers of photometric and spectroscopic redshifts, respectively, in parentheses.  Columns (4-8) indicate the number of sources that are individually detected in each waveband; fluxes from individual detections are combined with stacked fluxes of the remaining sources in each redshift bin, as discussed in the text and Appendix.}

\label{TABLE:RedshiftBinnedSample}
\end{deluxetable} 

\clearpage

%
%
%
\begin{deluxetable}{ccccccccccccc} 
\tablecolumns{11} 
\tablewidth{0pc} 
\tablecaption{UV-Radio Average Flux Densities for Redshift Binned s{\it BzK}s} 
\tablehead{ 
\colhead{ } &
\colhead{ } &
\multicolumn{2}{c}{$1.2<z\leq1.5$} &
\colhead{} &
\multicolumn{2}{c}{$1.5<z\leq2.0$} &
\colhead{} &
\multicolumn{2}{c}{$2.0<z\leq3.2$} \\
\colhead{Band}  & 
\colhead{$\lambda$} & 
\colhead{S$_\nu$}  &
\colhead{$\sigma_S$} &
\colhead{ } &
\colhead{S$_\nu$}  &
\colhead{$\sigma_S$} &
\colhead{ } &
\colhead{S$_\nu$}  &
\colhead{$\sigma_S$} \\
\colhead{(1)} &
\colhead{(2)} & 
\colhead{(3)} & 
\colhead{(4)} &
\colhead{ } & 
\colhead{(5)} & 
\colhead{(6)} &
\colhead{ } &
\colhead{(7)} & 
\colhead{(8)} &
 }

\startdata 
U          &      0.35 &      0.45 &      0.06 &&      0.36 &      0.05 &&      0.27 &      0.05      \\
B          &      0.46 &      0.61 &      0.04 &&      0.58 &      0.04 &&      0.59 &      0.04      \\
V          &      0.54 &      0.65 &      0.04 &&      0.67 &      0.04 &&      0.73 &      0.05      \\
R          &      0.65 &      0.85 &      0.05 &&      0.80 &      0.05 &&      0.86 &      0.05      \\
I          &      0.86 &      1.57 &      0.17 &&      1.36 &      0.16 &&      1.16 &      0.16      \\
z          &      0.90 &      1.63 &      0.24 &&      1.34 &      0.23 &&      1.18 &      0.24      \\
J          &      1.25 &      3.77 &      0.63 &&      4.31 &      0.63 &&      2.79 &      0.61      \\
H          &      1.65 &      4.25 &      0.79 &&      4.97 &      0.80 &&      3.52 &      0.71      \\
K          &      2.13 &      9.70 &      1.24 &&     10.32 &      1.22 &&      9.69 &      1.24      \\
24 $\mu$m  &     	24 &        94 &         2 &&       110 &      2.0 &&       120 &      2.1      \\
70 $\mu$m  &     	70 &       320 &        81 &&       340 &        83 &&       440 &        85      \\
250 $\mu$m &       250 &      2000 &       880 &&      9080 &       750 &&      1400 &       930      \\
350 $\mu$m &       350 &      2020 &       690 &&      5700 &       590 &&       920 &       730      \\
500 $\mu$m &       500 &      1030 &       480 &&      4300 &       400 &&      1300 &       510      \\
870 $\mu$m &       870 &       240 &        94 &&       510 &        80 &&       530 &       100      \\
1.4 GHz    &    214000 &      15 &       0.8 &&      12 &       0.7 &&      15 &       0.8      \\
610 MHz    &    490000 &      27 &       5.2 &&      32 &       4.3 &&      15 &       5.3 
\enddata 
\tablecomments{Column (1) $-$ waveband.  Column (2) $-$ effective, observed frame wavelength in units of $\mu$m.  Columns (3,5,7,9) $-$ average, observed flux density in units of $\mu$Jy for each redshift bin. Column (4,6,8,10) $-$ error in flux density in units of $\mu$Jy for each redshift bin.}
\label{Table:FluxDensity}
\end{deluxetable} 

\clearpage

%
%
%
%
%
\begin{deluxetable}{lllll}
\tablecolumns{5}
\tablecaption{Redshift Binned s{\it BzK} IR-Radio Fit Summary} 
\tablehead{
\colhead{Redshift} &
\colhead{Fit Type} &
\colhead{L$_{IR}$} &
\colhead{$\chi^2$(df) } &
\colhead{SFR$_{IR}$} \\
\colhead{(1)} & \colhead{(2)}  & \colhead{(3)}   & \colhead{(4)} &  \colhead{(5)} 
}
\startdata
  $1.2<z\leq 1.5$ & CE01 ($\ge$24 $\mu$m) & 3.0$\pm$0.3 &   48.23(7) & 51$\pm$5 \\
                  & CE01 ($>$24 $\mu$m) & 1.9$\pm$0.2 &    5.92(6) & 33$\pm$3 \\
                  & CE01 (24 $\mu$m) & 4.1$\pm$0.3 &    3.28(0) & 70$\pm$9 \\
  \cline{1-5}                
  $1.5<z\leq 2.0$ & CE01 ($\ge$24 $\mu$m) & 4.0$\pm$0.4 &  100.94(7) & 68$\pm$7 \\
                  & CE01 ($>$24 $\mu$m) & 5.0$\pm$0.9 &   75.31(6) & 86$\pm$15 \\
                  & CE01 (24 $\mu$m) & 8.6$\pm$0.5 &    0.03(0) & 148$\pm$45 \\
 \cline{1-5}                         
   $2.0<z\leq 3.2$ & CE01 ($\ge$24 $\mu$m) & 8.3$\pm$0.7 &   51.24(7) & 143$\pm$12 \\
                  & CE01 ($>$24 $\mu$m) & 6.3$\pm$0.7 &   13.23(6) & 108$\pm$12 \\
                  & CE01 (24 $\mu$m) & 35.4$\pm$2.1 &    1.82(0) & 608$\pm$91 \\
\enddata
\tablecomments{Column (1) indicates redshift range for the sample.  Column (2) specifies the type of model fit (see text).  Column (3) indicates the computed L$_{IR}$ from the best fit model, in units of 10$^{11}$ L$_\odot$.  Column (4) indicates the $\chi^2$ value for the best fit, with degrees of freedom in parentheses.  Column (5) indicates the SFR, in units of M$_\odot$ yr$^{-1}$, computed from the fit-derived IR luminosity.  Random errors correspond to 68\% confidence intervals, and do not include substantial systematic error, as discussed in the text.}
\label{TABLE:IRFITSUMMARY}
\end{deluxetable}

\clearpage

%
%
%
%
%
\begin{deluxetable}{rccccccc} 
\tablecolumns{8} 
\tablewidth{0pc} 
\tablecaption{Redshift Binned s{\it BzK} UV Luminosity and SFR Estimates} 
\tablehead{ 
\colhead{} & \multicolumn{5}{c}{Average of Individual Spectra} &   \colhead{}   &\colhead{Average Spectrum} \\ 
\cline{2-6} \cline{8-8} \\ 
\colhead{Redshift} & \colhead{$\beta_{fit}$}    &\colhead{log(IRX)} &		 \colhead{A$_{1600}$} &  \colhead{SFR$_{UV}^{Uncorr}$}   & \colhead{SFR$_{UV}^{Corr}$} & \colhead{} & \colhead{SFR$_{UV}^{Corr}$} \\
\colhead{(1)} & \colhead{(2)}  & \colhead{(3)}   & \colhead{(4)} &  \colhead{(5)}   & \colhead{(6)}   & \colhead{} & \colhead{(7)}  
}

\startdata
$1.2<z\leq1.5$ &     -1.50 &	1.12 &	1.44 &	4$\pm$0.1 &	12$\pm$1     	& &  10$\pm$3	\\ 
$1.5<z\leq2.0$ &     -0.80 &	2.15 &	2.84 &	6$\pm$0.3 &	64$\pm$4  	& & 36$\pm$12	\\ 
$2.0<z\leq3.2$ &     -0.59 &	2.21 &	3.26 &	11$\pm$0.8 &	285$\pm$30	& & 65$\pm$28	\\ 
\enddata

\tablecomments{Column (1) indicates sample redshift range.  Columns (2-6) correspond to unweighted averages of results from fitting individual galaxies in each redshift bin.  Column (2) indicates the best fit slope, $\beta$, to the $f_\lambda$ spectrum where $f_\lambda \propto \lambda^\beta$.  Column (3) indicates the IR-UV ratio, log(S$_{IR}$/S$_{1600}$).  Column (4) indicates the attenuation in magnitudes at 1600 \Angstrom derived from the best fit spectral index, $\beta$.  Column (5) and (6) refer to SFRs in units of M$_\odot$ yr$^{-1}$.  Column (7) indicates the SFR from IRX-$\beta$ correction applied to the corresponding (unweighted) redshift bin averaged spectrum.}
\label{Table:UVSFR}
\end{deluxetable} 

\clearpage

%
%
%
%
%
\begin{deluxetable}{llllll}
\tablecolumns{6}
\tablewidth{0pc} 
\tablecaption{Redshift Binned s{\it BzK} Radio Flux, Luminosity and SFR Estimates} 
\tablehead{
\colhead{Redshift} &
\colhead{S$_{1.4}$} &
\colhead{S$_{0.610}$} &
\colhead{L$_{1.4} \times$ 10$^{22}$} &
\colhead{SFR$_{1.4}^{Condon}$} &
\colhead{SFR$_{1.4}^{Bell}$} \\
\colhead{Range} &
\colhead{$\mu$Jy} &
\colhead{$\mu$Jy} &
\colhead{W Hz$^{-1}$} &
\colhead{M$_\odot$ yr$^{-1}$} &
\colhead{M$_\odot$ yr$^{-1}$} \\
\colhead{(1)}&
\colhead{(2)} & 
\colhead{(3)} & 
\colhead{(4)} & 
\colhead{(5)} & 
\colhead{(6)} 
}
\startdata
1.2$<$z$\leq$1.5 &   14.6 (12.1)$\pm$0.8 & 26.6 (14.8)$\pm$5.2 &    13.9 (11.5)$\pm$0.8 &  166 (138)$\pm$11 &   76 (64)$\pm$5 \\
1.5$<$z$\leq$2.0 &   11.7 (8.9)$\pm$ 0.7 &  32.2 (33.2)$\pm$4.3 &   19.4 (14.7)$\pm$1.1 &  232 (176)$\pm$16 &  107 (81)$\pm$7 \\
2.0$<$z$\leq$3.3 &   15.4 (8.4)$\pm$ 0.8 &  14.6 (17.2)$\pm$5.3 &   46.6 (25.3)$\pm$2.5 &  560 (303)$\pm$36 &  257 (139)$\pm$17 \\
\enddata
\tablecomments{Column (1) indicates the sample bin redshift range.  Columns (2) and (3) include individual detections and weighted average stacked flux at 1.4 GHz and 610 MHz respectively. In Columns (2,3,4-6), results of median stack of all sources are indicated in parentheses.  Column (4) indicates the derived rest frame 1.4 GHz luminosity.  Column (5) indicates the computed SFR according to \citet{1992ARA&A..30..575C}.  Column (6) indicates the computed SFR according to \citet{2003ApJ...586..794B}.}
\label{TABLE:RADIOSFR}
\end{deluxetable}

\clearpage

%
%
%
\begin{deluxetable}{rcccccc}
\tablecolumns{7}
\tablewidth{0pc}
\tablecaption{Redshift Binned s{\it BzK} X-ray Flux, Luminosity and SFR Estimates}
\tablehead{
\colhead{Redshift} & \colhead{Flux}   & \colhead{Flux}    &\colhead{Luminosity} &    \colhead{SFR$^{Ranalli}_{2-10 keV}$} & \colhead{SFR$^{Persic}_{2-10 keV}$}    & \colhead{SFR$^{Lehmer}_{2-10 keV}$} \\
\colhead{Range} & \colhead{0.5-2 keV}   & \colhead{2-8 keV}    &\colhead{Rest 2-10 keV} &    \colhead{$M_\odot$ yr$^{-1}$} & \colhead{$M_\odot$ yr$^{-1}$}    & \colhead{$M_\odot$ yr$^{-1}$} \\
\colhead{(1)} & \colhead{(2)}  & \colhead{(3)}   & \colhead{(4)} &  \colhead{(5)}   & \colhead{(6)}   & \colhead{(7)}
}
\startdata
$1.2 < z \leq 1.5$ &   3.2$\pm$1.0  &  31$\pm$21  &            7$\pm$7  &           15$\pm$22  &           73$\pm$67  & 62$\pm$46 \\
$1.5 < z \leq 2.0$ &   7.4$\pm$1.2  &   44$\pm$16 &           29$\pm$10  &           58$\pm$75  &          292$\pm$116  &  291$\pm$123 \\
$2.0 < z \leq 3.2$ &   9.5$\pm$1.2  &  27$\pm$ 6   &           50$\pm$14  &          100$\pm$126  &          499$\pm$170  & 507$\pm$203 \\
\enddata
\tablecomments{Column (1) $-$ the sample bin redshift range. Column (2) $-$ observed flux density in {\it Chandra} soft band (0.5-2 keV), in units of 10$^{-18}$ erg s$^{-1}$ cm$^{-2}$.  Column (3) $-$  observed flux density in {\it Chandra} hard band (2-8 keV), in units of 10$^{-18}$ erg s$^{-1}$ cm$^{-2}$ Column (4)  $-$ the luminosity in rest frame (2-10 keV), computed from observed, soft band flux, assuming spectrum with photon index, $\Gamma$=1.2 and E$_c$=20 keV, at the median redshift.  Units are 10$^{40}$ erg s$^{-1}$.  Column (5) $-$ SFR from the method of \citet{2003A&A...399...39R}.  Column (6) $-$ SFR from the method of \citet{2004A&A...419..849P}.  Column (7) $-$ SFR from the method of \citet{2010ApJ...724..559L}.  All SFRs in units of M$_\odot$ yr$^{-1}$ assuming Salpeter IMF (0.1 - 100 M$_\odot$).}
\label{TABLE:XRAYSFR}
\end{deluxetable}

\clearpage

%
%
%
%
%
\begin{turnpage}
\begin{deluxetable}{rccccccccc} 
\tablecolumns{10} 
\tablewidth{0pc} 
\tablecaption{Redshift Binned s{\it BzK} Star Formation Rate Summary} 
\tablehead{ 
\colhead{Redshift} & \colhead{SFR$_{IR}$}   & \colhead{SFR$_{24 \mu m}$}    & \colhead{SFR$_{UV}^{Corr}$} & \colhead{SFR$_{1.4 GHz}^{Condon}$} & \colhead{SFR$_{1.4 GHz}^{Bell}$} & \colhead{SFR$^{Ranalli}_{2-10 keV}$} & \colhead{SFR$^{Persic}_{2-10 keV}$} & \colhead{SFR$^{Lehmer}_{2-10 keV}$}   & \colhead{SFR$_{IR+UV}$}\\
\colhead{(1)} & \colhead{(2)} & \colhead{(3)} & \colhead{(4)} & \colhead{(5)} & \colhead{(6)} & \colhead{(7)} &\colhead{(8)}& \colhead{(9)} &\colhead{(10)} }

\startdata 
$1.2<z\leq1.5$          & 	{\bf 51$\pm$5}     		& {\bf 79$\pm$9} 	&    	  12$\pm$1 	& 138$\pm$11 		& {\bf 64$\pm$5}		& 15$\pm$22 	         & 73$\pm$67	       	& 62$\pm$46  		& {\bf 55$\pm$6}   \\
$1.5<z\leq2.0$          & 	{\bf 68$\pm$7}      		& {\bf 172$\pm$51}	&    	{\bf 64$\pm$4} 	& 176$\pm$16		& {\bf 81$\pm$7}		& {\bf 58$\pm$75}	& 292$\pm$116  	& 291$\pm$123	& {\bf 74$\pm$8}   \\
$2.0<z\leq3.2$          &       {\bf 143$\pm$12}   	   	& 620$\pm$85		&    	 285$\pm$30	& 303$\pm$36 		& {\bf 139$\pm$17}		& 100$\pm$126      	& 499$\pm$170     	& 507$\pm$203	& {\bf 154$\pm$17} \\
\enddata 
\tablecomments{Star formation rates for redshift binned s{\it BzK}s.  Column (1) $-$ sample redshift range.  Column (2) $-$ SFR  from integrated IR luminosity using CE01 template fits to MIR-radio photometry ($\lambda \geq 24 \mu$m).  Column (3) $-$ SFR from CE01 template fit to 24 $\mu$m data only. Column (4) $-$ SFR from average of individual fits to UV continuum with dust correction.   Columns (5) and (6) $-$ SFR  from median, stacked 1.4 GHz flux using the method of  \citet{1992ARA&A..30..575C} and \citet{2003ApJ...586..794B} respectively.  Columns (7-9) $-$ SFR from soft band X-ray data using the method of \citet{2003A&A...399...39R}, \citet{2004A&A...419..849P}  and \citet{2010ApJ...724..559L} respectively.  Column (10) $-$ the total SFR from the sum of Column (2) and corresponding uncorrected UV SFRs taken from Table \ref{Table:UVSFR}.  Units are M$_\odot$ yr$^{-1}$ for all columns.  Values in boldface fall within 2$\sigma$ of the best estimate, SFR$_{IR+UV}$.  Errors do not include substantial systematic uncertainty, as discussed in the text.}
\label{Table:SFR}
\end{deluxetable} 
\end{turnpage}

\end{document}